\documentclass[
superscriptaddress,
nofootinbib,
twocolumn,
 amsmath,amssymb,
 aps,
pra,
notitlepage,
longbibliography
]{revtex4-1}

\usepackage{dcolumn}% Align table columns on decimal point
\usepackage{bm}% bold math
\usepackage{epsfig}
\usepackage{graphicx}
\usepackage{latexsym}
\usepackage{amsfonts}
\usepackage{setspace}
\usepackage{graphicx}
\usepackage{amsmath}
\usepackage{mathrsfs}
\usepackage{verbatim}
\usepackage{color}
\usepackage{SIunits}
\usepackage{hyperref}
\usepackage{physics}
\newcommand{\be}{\begin{equation}}
\newcommand{\ee}{\end{equation}}
\newcommand{\ba}{\begin{array}}
\newcommand{\ea}{\end{array}}
\newcommand{\bqa}{\begin{eqnarray}}
\newcommand{\eqa}{\end{eqnarray}}

\newcommand{\np}[1]{\textcolor{black}{#1}}

\begin{document}

\title{\np{Quantum correlated photons via a passive nonlinear microcavity}}

\author{Mengdi Zhao} 
%\thanks{These authors contributed equally to this work.}
\affiliation{Holonyak Micro and Nanotechnology Laboratory, University of Illinois at Urbana-Champaign, Urbana, IL 61801 USA}
\affiliation{Department of Physics, University of Illinois at Urbana-Champaign, Urbana, IL 61801 USA}
\affiliation{Illinois Quantum Information Science and Technology Center, University of Illinois at Urbana-Champaign, Urbana, IL 61801 USA}
\author{Yunkai Wang} 
%\thanks{These authors contributed equally to this work.}
\affiliation{Holonyak Micro and Nanotechnology Laboratory, University of Illinois at Urbana-Champaign, Urbana, IL 61801 USA}
\affiliation{Department of Physics, University of Illinois at Urbana-Champaign, Urbana, IL 61801 USA}
\affiliation{Illinois Quantum Information Science and Technology Center, University of Illinois at Urbana-Champaign, Urbana, IL 61801 USA}
\author{Shanhui Fan} 
%\thanks{These authors contributed equally to this work.}
\affiliation{Department of Electrical Engineering, Ginzton Laboratory, Stanford University, Stanford, CA 94305, USA}
\author{Kejie Fang} 
\email{kfang3@illinois.edu}
%\homepage{https://fang.ece.illinois.edu}
\affiliation{Holonyak Micro and Nanotechnology Laboratory, University of Illinois at Urbana-Champaign, Urbana, IL 61801 USA}
\affiliation{Illinois Quantum Information Science and Technology Center, University of Illinois at Urbana-Champaign, Urbana, IL 61801 USA}
\affiliation{Department of Electrical and Computer Engineering, University of Illinois at Urbana-Champaign, Urbana, IL 61801 USA}

\begin{abstract} 

Photons, by nature, typically do not exhibit interactions with each other. Creating photon-photon interactions holds immense importance in both fundamental physics and quantum technologies. Currently, such interactions have only been achieved indirectly as mediated by atomic-like quantum emitters with resonant photon-atom interactions. However, the use of these indirect interactions presents substantial fundamental challenges that impede scaling and practical applications. \np{Here we demonstrate creation of non-classical photon correlations, including photon anti-bunching, via a passive InGaP photonic integrated circuit. Our approach employs the quantum interference between uncorrelated light and the two-photon bound state, the latter of which arises from the $\chi^{(2)}$-mediated photon interaction.} Our work opens a new route in controlling quantum light by harnessing highly-engineerable bulk optical nonlinearities, which has significant implications for nonlinear optical quantum information processing and quantum networking.

\end{abstract}
%\pacs{}

\maketitle

Interactions between single photons form the foundation of gate-based optical quantum computing and can enable numerous quantum technologies \cite{chang2014quantum}. Such interactions, manifested in a correlated two-photon scattering process \cite{birnbaum2005photon,faraon2008coherent,reinhard2012strongly,firstenberg2013attractive,cantu2020repulsive,prasad2020correlating,le2021experimental,masters2023simultaneous}, have only been achieved indirectly using atomic-like quantum emitters through resonant photon-atom interactions. Often, these emitters are embedded in optical cavities to further enhance photon-atom interactions. However, the use of such indirect interactions poses substantial fundamental challenges that limit scaling and practical applications.
For example, in these systems, the operating wavelength and bandwidth are constrained by the intrinsic properties of the quantum emitters and cannot be arbitrarily chosen. Atomic-like emitters are also subject to increased thermal fluctuations, necessitating laser or passive cooling. Furthermore, in the case of solid-state quantum emitters, each individual emitter can be in a different microscopic environment, leading to substantial inhomogeneity in the resonant wavelengths. Such inhomogeneity poses a significant challenge for scaling up quantum circuits based on these emitters.

The challenges associated with quantum emitters can be overcome by employing direct photon-photon interactions \cite{munro2005weak,langford2011efficient}. 
Direct photon-photon interactions make use of non-resonant nonlinearities, such as $\chi^{(2)}$ or $\chi^{(3)}$ processes, in bulk transparent optical materials. The utilization of direct photon-photon interactions via bulk nonlinearities holds great promise for numerous schemes in nonlinear optical quantum information processing \cite{munro2005weak, langford2011efficient,niu2018qudit,li2020photon,krastanov2021room,pick2021boosting,yanagimoto2022temporal}. Bulk nonlinear processes have also been extensively explored in the field of quantum optics, including for generation of non-classical photon correlations \cite{walborn2010spatial,majumdar2013single,flayac2017unconventional}. Recent experiments have demonstrated three-photon correlations via cascaded frequency-up or -down conversion processes \cite{hubel2010direct,guerreiro2013interaction}. However, \np{direct generation of non-classical correlations upon flying uncorrelated photons via passive bulk nonlinearities} remains an outstanding challenge, as the required strength of nonlinearity appears to surpass current experimental capabilities \cite{langford2011efficient,hubel2010direct,guerreiro2013interaction}. 

In this study, we experimentally generated single-mode non-classical correlations of light induced by direct photon interactions in a nonlinear monolithic photonic integrated circuit. Our approach utilizes the material platform of InGaP, which has a strong $\chi^{(2)}$ nonlinearity with low losses, and moreover exploits the dissipation engineering technique \cite{harrington2022engineered} to manipulate the photon correlations. By manipulating the quantum interference between the unscattered photons and $\chi^{(2)}$-induced two-photon bound state \cite{shen2007strongly} through a waveguide-coupled nonlinear microcavity, we create strong quantum correlations between transported photons, including photon anti-bunching. Our findings opens up possibilities for generation of strongly correlated multi-photon states in integrated nonlinear photonic platforms for quantum information processing and quantum simulation \cite{slussarenko2019photonic,altman2021quantum}.

\begin{figure*}[!htb]
\begin{center}
\includegraphics[width=2\columnwidth]{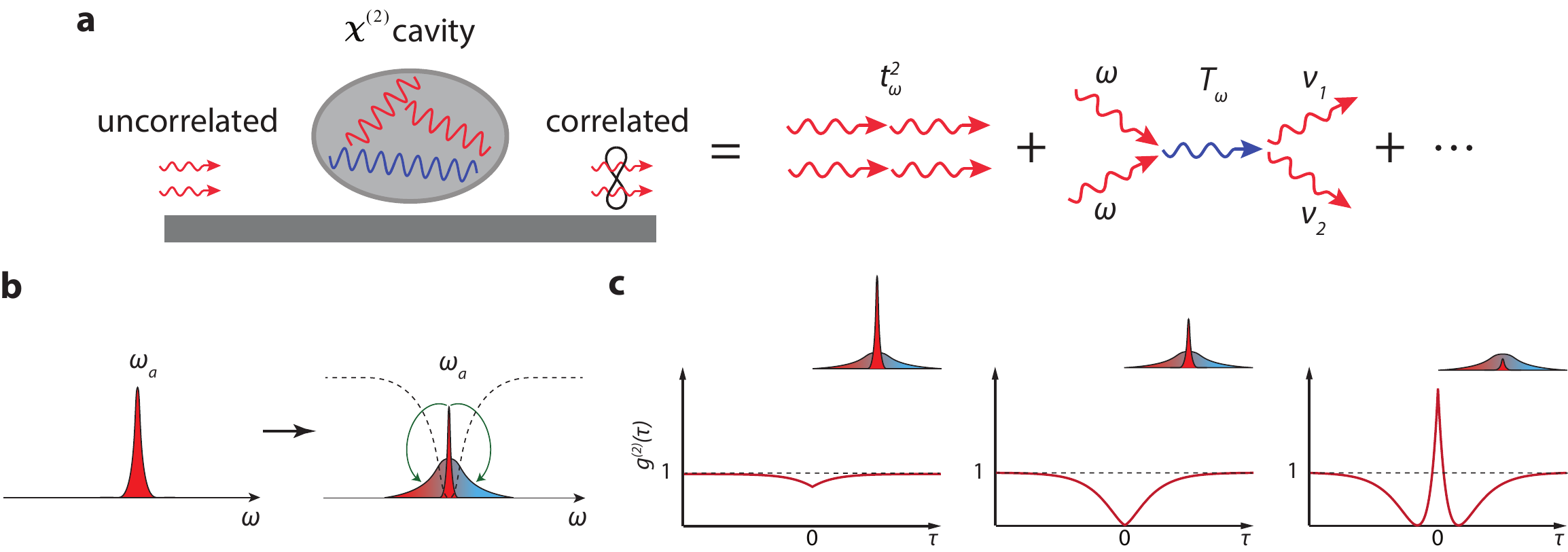}
\caption{\textbf{Two-photon transport via a waveguide-coupled $\chi^{(2)}$ cavity.} \textbf{a}. Diagrammatic representation of the two-photon transport via a waveguide-coupled $\chi^{(2)}$ cavity, involving quantum interference between the unscattered transport and second-harmonic photon-mediated scattering. \textbf{b}. The $\chi^{(2)}$ interaction causes scattering of input photons (red envelope), resulting in correlated photons and the two-photon bound state with an expanded spectrum (gradient-color envelope) modulated by the cavity resonance (dashed line). \textbf{c}. Second-order correlation function of the transported photons depending on the ratio of the bound state and unscattered photon amplitude. The latter is controlled by the linear transmission coefficient $t_{\omega}$.}
\label{fig:scheme}
\end{center}
\end{figure*}

Our experimental system is schematically illustrated in Fig. \ref{fig:scheme}a, which includes a $\chi^{(2)}$ nonlinear cavity coupled to a waveguide. Photon interactions via the nonlinear cavity can alter the correlations of the light transmitted through the waveguide.
For a doubly-resonant $\chi^{(2)}$ nonlinear cavity, the interaction Hamiltonian is given by
\be\label{Hchi2}
\hat H_{\textrm{int}}= \hbar g(\hat a^{\dagger 2}\hat b+\hat a^2\hat b^\dagger),
\ee
where $\hat a(\hat a^\dagger)$ and $\hat b(\hat b^\dagger)$ are the annihilation (creation) operators for the fundamental and second-harmonic resonances with frequencies $\omega_b\approx2\omega_a$, respectively, and $g$ is the single-photon nonlinear coupling.  

\np{Few-photon transport via a waveguide-coupled nonlinear quantum system have been  explored widely in the context of waveguide quantum electrodynamics (QED) using quantum emitters \cite{shen2007strongly, chang2007single, sipahigil2016integrated,foster2019tunable, prasad2020correlating, le2021experimental}. However, similar effects have not been studied for a nonlinear cavity, given the lack of sufficient bulk optical nonlinearity.} Using the full quantum mechanical $S-$matrix formalism \cite{wang2022few,xu2015input,xu2013analytic}, the two-photon scattering amplitude through the waveguide, $S_{\nu_1\nu_2;\omega_1\omega_2}\equiv{}_{\textrm{out}}\langle \nu_1\nu_2|\omega_1\omega_2\rangle_{\textrm{in}}$, can be expressed as \cite{xu2013analytic}
\begin{equation}\label{Smatrix}
S_{\nu_1\nu_2;\omega_1\omega_2}=S_{\nu_1\nu_2;\omega_1\omega_2}^0+iT_{\nu_1\nu_2;\omega_1\omega_2},
\end{equation}
where $\omega_i$ and $\nu_i$ represent the (angular) frequency of the incoming and outgoing photons in the waveguide, respectively. The uncorrelated transport amplitude, $S_{\nu_1\nu_2;\omega_1\omega_2}^0$, is given by the product of single-photon transmission coefficients of the two photons, while the interaction-induced, correlated transport amplitude is given by $iT_{\nu_1\nu_2;\omega_1\omega_2}$ (see Methods). Fig. \ref{fig:scheme}a illustrates this process for a $\chi^{(2)}$ cavity, where the correlated transport, to the leading order, is mediated by a second-harmonic photon in the cavity. During this process, the two incoming photons exchange energy via the intermediate photon while preserving the total energy, i.e., $\hbar \nu_1+\hbar \nu_2=\hbar \omega_1+\hbar \omega_2$ (for arbitrary $\nu_{1(2)}$), leading to the two-photon bound state with an energy of $\hbar \omega_1+\hbar \omega_2$ \cite{shen2007strongly,firstenberg2013attractive,javadi2015single}. \np{The two-photon bound state thus has an expanded spectrum modulated by the cavity bandwidth (Fig. \ref{fig:scheme}b). The two-photon bound state, together with the unscattered photons, constitutes the complete basis of the Hilbert space of the asymptotic two-photon states in the waveguide-coupled nonlinear system \cite{shen2007strongly}, regardless of the strength of the photon interaction, and thus can be observed even in the weak coupling regime \cite{prasad2020correlating}. }

\begin{figure*}[!htb]
\begin{center}
\includegraphics[width=2\columnwidth]{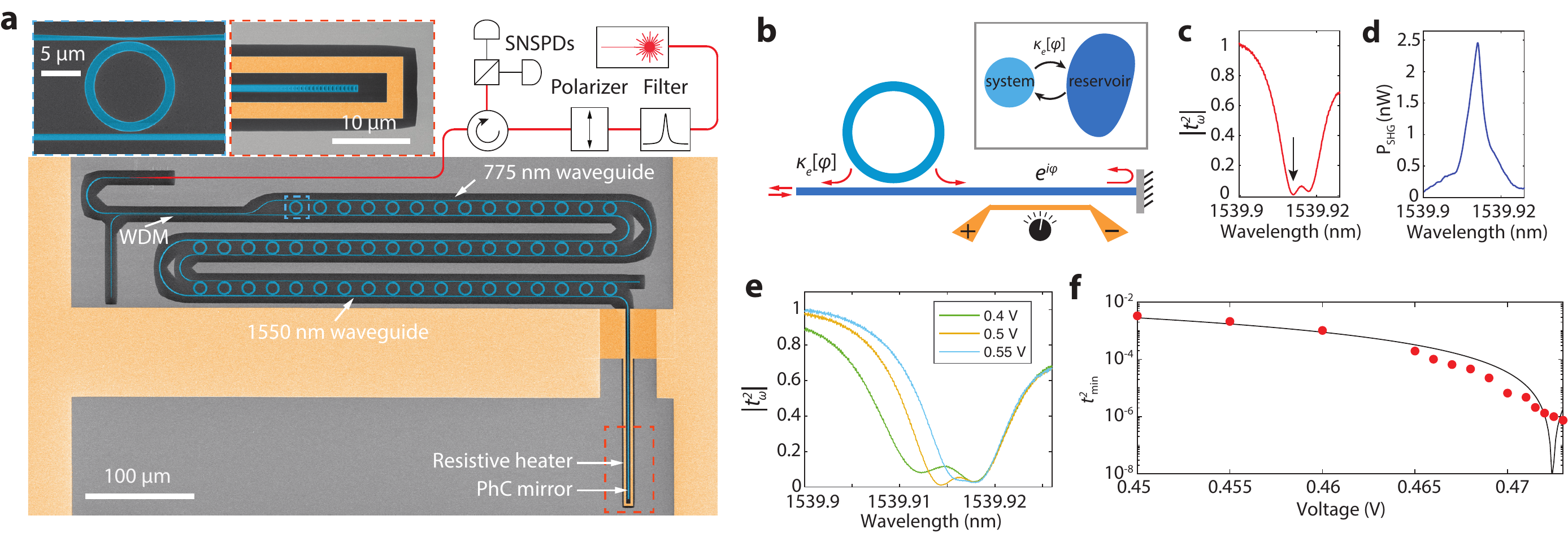}
\caption{\textbf{Dissipation-engineered nonlinear photonic circuit.} \textbf{a}. False-color scanning electron microscope images of the fabricated photonic circuit. Insets show the microring resonator and tunable waveguide with a resistive heater. WDM: wavelength-division multiplexer. SNSPD: superconducting nanowire single-photon detector. \textbf{b}. Dissipation engineering of the microring resonator via the waveguide feedback. The external cavity-photon dissipation is controlled by the waveguide phase $\varphi$. \textbf{c}. Normalized transmission spectrum of the phase- and frequency-matched resonance (arrow) used in the experiment. \textbf{d}. Second-harmonic power spectrum corresponding to $P_{\rm in}=1.49~\mu$W. \textbf{e}. Tuning of the transmission spectrum of the phase-matched resonance for various applied voltages.  \textbf{f}. Fine-tuning of the minimum transmission at 1539.914 nm. Solid line is the model fitting (SI).}
\label{fig:setup}
\end{center}
\end{figure*}

The second-order correlation function of transported photons for monochromatic input photons with a frequency $\omega$ is related to the $S$-matrix and can be expressed as (see Methods)
\be\label{g2}
g^{(2)}(\tau)\equiv\frac{|\psi_2(t,t+\tau)|^2}{|\psi_1(t)|^4}=\frac{|t_\omega^2+ T(\omega,\tau)|^2}{|t_\omega^2|^2},
\ee 
where $\psi_2(t,t+\tau)$ and $\psi_1(t)$ denote the temporal two-photon and single-photon wavefunction, respectively, and $t_\omega$ is the single-photon transmission coefficient.
$T(\omega,\tau)$, given by Eq. \ref{T}, corresponds to the wavefunction of the two-photon bound state, which has an amplitude proportional to $g^2/\kappa_a\kappa_b$ and a temporal extent of $1/\kappa_a$ \cite{shen2007strongly}, where $\kappa_{a(b)}$ is the photon dissipation rate of resonance $a(b)$. The correlation of the transported photons thus is determined by the quantum interference between the unscattered transport and the two-photon bound state (Fig. \ref{fig:scheme}c) \cite{wang2022few}. We can generate strongly correlated photons by controlling the single-photon transmission of the photonic circuit and tune the system across an extremely rich space of accessible photon correlations by adjusting, for example, the frequency of the light and the dissipation of the cavity photon. 
Our approach is different from the so-called unconventional photon blockade \cite{flayac2017unconventional,snijders2018observation,vaneph2018observation}, which could be explained as an optimal squeezing effect \cite{lemonde2014antibunching} (see SI for more discussions).

Our experiment utilizes the $\chi^{(2)}$ InGaP photonic platform with a high nonlinearity-to-loss ratio \cite{zhao2022ingap}. The fabricated InGaP photonic circuit is shown in scanning electron microscope images in Fig. \ref{fig:setup}a (see Methods and SI for fabrication and device details). The nominal microring resonator is designed to support phase- and frequency-matched fundamental 1550-nm-band transverse-electric (TE$_{00}$) and second-harmonic 775-nm-band transverse-magnetic (TM$_{00}$) resonances for $\chi^{(2)}$-enabled photon interaction. 
In general, phase-matching condition places a stringent requirement on the device parameters. Since parameter variations due to fabrications are unavoidable, we utilize a photonic circuit consisting of multiple microrings, each of which has a slightly different ring width, to ensure that one of these microrings can satisfy the phase-matching condition. In this circuit, the nonlinear interaction occurs only in the single microring that satisfies the phase-matching condition. The strong photon confinement in such a microring enhances the nonlinear interaction.  We also implement a mechanism to tune the cavity dissipation rate of the microring by placing a reflector at the end of one of the waveguides coupling to the ring to create a standing-wave resonator, as shown in Fig. \ref{fig:setup}b. In our experiment the reflector is implemented using a photonic crystal mirror (Fig. \ref{fig:setup}a)
The tunable cavity dissipation is approximately given by $\kappa_{ae}[\varphi]=\frac{1}{2}\kappa_{e0}(1-\sin2\varphi)$, where $\kappa_{e0}$ is the external cavity dissipation rate of the ring in the absence of the mirror and $\varphi$ is the waveguide phase between the microring and the photonic crystal mirror \cite{zhao2022observation} (SI). Via an integrated resistive heater, we can tune the waveguide phase $\varphi$ and therefore the single-photon transmission coefficient of the one-port photonic circuit, which depends on $\kappa_{ae}$. The waveguide feedback also induces a frequency detuning, $\Delta$, between the local minimum of the transmission spectrum and the bare cavity resonance.

\begin{figure*}[!htb]
\begin{center}
\includegraphics[width=2\columnwidth]{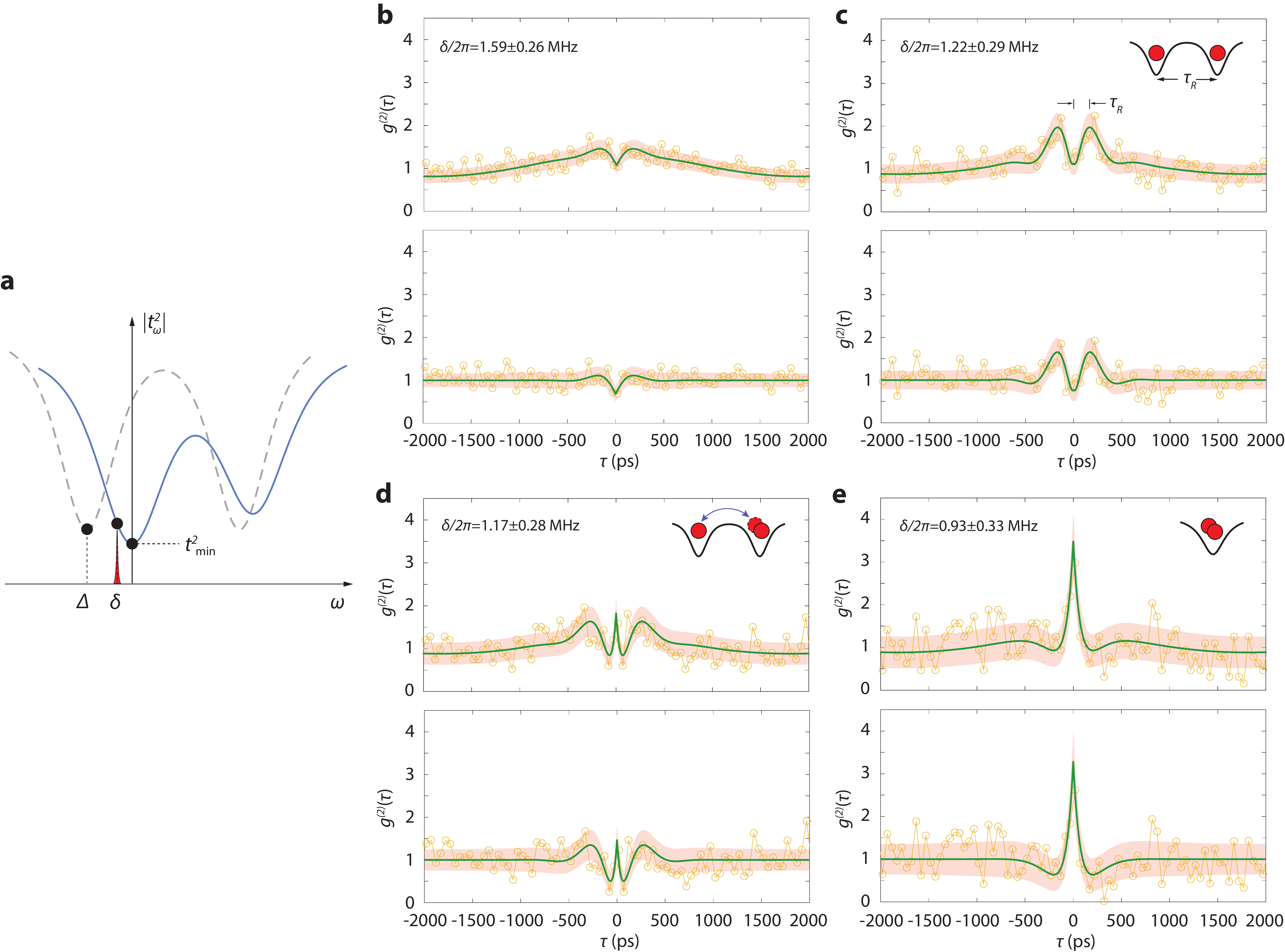}
\caption{\textbf{Tunable quantum correlations of photons.}  \textbf{a}. Schematic plot illustrates the parameters used to control $g^{(2)}(\tau)$. Blue and dashed lines represent the transmission spectrum $|t_\omega|^2$ and the bare cavity resonances, respectively. Laser detuning is $\delta$. \textbf{b}--\textbf{e}. Measured $g^{(2)}(\tau)$ of transported photons corresponding to different frequency detunings ($\delta/2\pi=1.59\pm0.26, 1.22\pm0.29, 1.17\pm0.28, 0.93\pm0.33$ MHz) and $P_{\rm in}=56.3$ nW. Top panels are raw $g^{(2)}(\tau)$ and bottom panels are background subtracted $g^{(2)}(\tau)$. \textbf{b} and \textbf{c} show the photon anti-bunching, i.e., $g^{(2)}(0)<g^{(2)}(\tau)$ for some $\tau$.  $g^{(2)}(\tau)$ is normalized with respect to $g^{(2)}(\tau=12.5~ \rm ns)$ (SI). Green line is model fitting of the data. The shaded area represents $\pm$ 1$\sigma$ deviation of the fitted curve.  See SI for fitting details. }
\label{fig:repulsion}
\end{center}
\end{figure*}
 
Fig. \ref{fig:setup}c shows the spectrum of a pair of split 1550-nm band TE$_{00}$ standing-wave resonances of a microring resonator. The intrinsic quality factors of the 1550-nm band TE$_{00}$ resonance and 775-nm band TM$_{00}$ resonance are typically about $Q_{ai}\approx 2.5\times 10^5$ and $Q_{bi}\approx 5\times 10^4$, respectively. To characterize the nonlinear mode coupling $g$, we used the process of second harmonic generation (SHG). The 1550-nm-band TE$_{00}$ resonance that satisfies the frequency- and phase-matching condition with a 775-nm-band TM$_{00}$ resonance of the same microring allows for strong SHG. Fig. \ref{fig:setup}d shows the measured second-harmonic signal from the phase-matched resonance around 1539.914 nm for an input laser power of $P_{\textrm{in}}=1.49\ \mu$W. The nonlinear mode coupling is extracted from the SHG efficiency to be $g/2\pi=6.5$ MHz (SI), resulting in a nonlinearity-to-loss ratio of $g/\kappa_{ai}=0.84\%$ or $g/\sqrt{\kappa_{ai}\kappa_{bi}}=0.27\%$.

The measurement of photon correlations is carried out at a temperature of 4 K in order to reduce the effects of thermomechanical noise originating from the vibrational modes of the membrane device and thermorefractive noise \cite{zhao2022observation}. Nevertheless, operating the device at room temperature is possible by utilizing the InGaP-on-insulator architecture \cite{martin2017gainp} or adding top oxide cladding on the microring, which will suppress the optomechanical coupling with the vibrational modes and temperature fluctuations (SI). Prior to coupling the laser into the device, it is filtered to eliminate amplified spontaneous emission, side-mode emission, and unpolarized light (Methods and SI). Correlations of the transmitted light are measured in a Hanbury Brown-Twiss setup. The phase-matched resonance at 1539.914 nm can be tuned from under-coupling ($\kappa_{ai}>\kappa_{ae}$) to over-coupling ($\kappa_{ai}<\kappa_{ae}$) with a tuning voltage of less than 1 V (Fig. \ref{fig:setup}e). In the vicinity of the critical-coupling regime ($V\approx 0.47$ V), the local minimum of the transmission spectrum, $t_{\mathrm{min}}^2$, can be precisely tuned with a resolution of about $10^{-6}$ using voltage increments of 1 mV (Fig. \ref{fig:setup}f). The minimum achievable value of $t_{\mathrm{min}}^2$ is limited by residual thermomechanical noise (SI).

We perform the measurement using a weak coherent light with wavelength around the local transmission minimum at 1539.914 nm as the input signal of the photonic circuit. The power of the coherent field satisfies the weak driving condition, resulting in single intermediate second-harmonic photons, i.e., $\bar n_{c,\textrm{SH}}\approx 4(g\bar n_c/\kappa_b)^2<1$. In our experiment, the on-chip input optical power is $P_\mathrm{in}=56.3$ nW, corresponding to 
a mean cavity second-harmonic photon number $\bar n_{c,\mathrm{SH}}=0.021$. The process of photon scattering mediated by the second-harmonic photon thus is in the single-photon nonlinear regime, where the parametric approximation of the nonlinear process is not valid \cite{crouch1988limitations,scharf1984effect}, and instead the full nonlinear Hamiltonian (Eq. \ref{Hchi2}) needs to be used.  In this regime, the two-photon component of the weak coherent state is subject to the $\chi^{(2)}$-enabled interaction in the cavity, leading to second-order quantum correlations of the transmitted light given by Eq. \ref{g2} (SI). The quantum correlation of transmitted photons can be controlled by the signal-local minimum detuning, $\delta$, and the waveguide phase, $\varphi$, which sets the local minimum transmission, $t^2_{\mathrm{min}}$, and the local minimum-cavity resonance detuning, $\Delta$ (Fig. \ref{fig:repulsion}a). 

We maintain the local minimum of the transmission spectrum within the range of 6--9$\times 10^{-7}$ and only vary the frequency of the input light to create various quantum correlations. The measured correlations $g^{(2)}(\tau)$ for single-photon transmission coefficients $|t_\omega|^2=(5.6\pm1.4)\times 10^{-6}$ and $(3.9\pm1.2)\times 10^{-6}$ are shown in the top panels of Fig. \ref{fig:repulsion}b and c, respectively. The distinct features of $g^{(2)}(\tau)$ in the range $|\tau|\lesssim2\pi/\kappa_a(\approx600$ ps) are due to the photon interaction in the $\chi^{(2)}$ cavity with a bandwidth of $\kappa_a/2\pi$($\approx 1.5$ GHz). The green curves in the plots represent the model fitting (SI). The background correlation is due to the mixture of the residual thermomechanical noise and the coherent photons  \cite{zhao2022observation}, which is universal for all ring resonances (see SI for the measured $g^{(2)}(\tau)$ of a phase-unmatched resonance). \np{$g^{(2)}(\tau)$ is normalized with respect to $g^{(2)}(\tau=12.5~ \rm ns)$ (SI), which is why $g^{(2)}(\tau=2000~ \rm ps)$ appears to be less than 1 for the fitted curve.}

\np{As seen from Figs. \ref{fig:repulsion}b and c, quantum correlations manifested as the anti-bunching effect, which is defined as having $g^{(2)}(0)<g^{(2)}(\tau)$ for some $\tau > 0$ \cite{scully1997quantum}, are observed for the transported photons without background subtraction (top panels), indicating a suppressed likelihood of finding two photons at $\tau=0$. In contrast, classical light satisfies the inequality $g^{(2)}(0)\ge g^{(2)}(\tau)$ for all $\tau$. Our data shows this inequality is strongly violated by 2.3$\sigma$ and 2.8$\sigma$ for the photon correlations of Figs. \ref{fig:repulsion}b and c, respectively, calculated as $g^{(2)}(\tau=170~ {\rm ps})-g^{(2)}(0)$, where $\sigma$ is one standard deviation of $g^{(2)}(0)$.} The cause of this anti-bunching effect is attributed to the swap between the two fundamental photons and the intermediate second-harmonic photon, which leads to a geometric $\pi$ phase shift of the scattered state and thus cancellation with the unscattered amplitude. \np{We stress photon anti-bunching should not be confused with sub-Poissonian statistics, i.e., $g^{(2)}(0)<1$ \cite{zou1990photon}. Eliminating the residual thermomechanical noise, e.g., via adding top oxide cladding of the microring to suppress the optomechanical coupling between the microring and membrane vibration modes \cite{zhao2022observation} (SI), will allow the observation of sub-Poissonian statistics of the transmitted coherent light (Fig. \ref{fig:scheme}c). To illustrate this, Figs. \ref{fig:repulsion}b and c bottom panels show background subtracted $g^{(2)}(\tau)$, which reveals that the coherent part of the transmitted light is indeed sub-Poissonian. }

The observed correlations shown in Figs. \ref{fig:repulsion}b and c (without background subtraction) also violate another classical inequality given by $|g^{(2)}(0)-1|\ge|g^{(2)}(\tau)-1|$ \cite{rice1988single}. In addition, because $g^{(2)}(\tau)$ is proportional to the two-photon probability distribution, Fig. \ref{fig:repulsion}c also indicates an effective photon repulsion effect, where a peaked correlation $g^{(2)}(\tau)>1$ at a finite delay $\tau=\tau_R (\approx 200$ ps) means that the two photons are pushed away from each other with a temporal separation $\tau_R$. The appearance of the peaked $g^{(2)}(\tau)$ at $\tau=\tau_R$ is related to the detuning between the signal and bare cavity resonance, $|\Delta-\delta|$, which leads to an oscillation of $T(\omega,\tau)$ with a period $2\pi/|\Delta-\delta|$ (see Eq. \ref{T}) and thus quantum beat of the scattered and unscattered photon components \cite{legero2004quantum}. In contrast, the interference in the linear photonic circuit alone is not able to produce such single-mode correlation features. It is necessary to have $\Delta\gtrsim \kappa_a$ in order to observe such quantum beat since the bound-state amplitude $T(\omega,\tau)$ has a decay time of $1/\kappa_a$. The photonic circuit architecture with split standing-wave resonances and tunable waveguide feedback makes this condition possible. For the phase-matched resonance here, we found $\Delta/2\pi\approx 1.6$ GHz from model fitting and $|\delta|/2\pi\approx$ 1--3 MHz. 

When the input photon frequency approaches the local minimum of the transmission spectrum, the two-photon wavefunction becomes dominated by the $\chi^{(2)}$ interaction-induced two-photon bound state amplitude $T(\omega,\tau)$, as shown in Fig. \ref{fig:repulsion}d and e. Fig. \ref{fig:repulsion}d shows the measured correlation $g^{(2)}(\tau)$ at $|t_\omega|^2=(3.7\pm1.1)\times 10^{-6}$. It exhibits positive correlations ($g^{(2)}>1$) at both zero delay and some finite delay, with anti-correlation ($g^{(2)}<1$) between the two $g^{(2)}$ peaks. Fig. \ref{fig:repulsion}e shows the dominant bound state at $|t_\omega|^2=(2.9\pm1.0)\times 10^{-6}$ characterized by $g^{(2)}(0)=2.98\pm0.68$ and $g^{(2)}(0)>g^{(2)}(\tau)$, meaning the two photons are attracted to each other with increased likelihood of being found together.  \np{The output photon flux from the device in this case is $1.6\times 10^6$ photons/s, corresponding to a photon number of $3.3\times 10^{-4}$ within the bandwidth of the resonator. As a result, multiphoton components can be ignored in the output state and the bunching is due to the two-photon bound state. Using the measured $g^{(2)}(0)$, the two-photon bound state amplitude in the transmitted light is estimated to be $(64\pm8)\%$ according to Eq. \ref{g2}.} 

Our experimental findings present a novel method for generating strongly-correlated photons in nonlinear optical systems without requiring strong light-matter interaction
or quantum emitters. The use of material’s intrinsic nonlinearity allows a large degree of control over the photon correlation in a scalable integrated photonic platform. Compared with atomic-like systems, the nonlinear cavity utilized in our scheme possesses a significantly larger mass, resulting in reduced thermal fluctuations, and thus can in principle be operated at room temperature. Moreover, for an on-chip optical cavity, the operating wavelength and bandwidth are controlled entirely by the properties of the cavity and the cavity-waveguide coupling, which can be defined lithographically and arbitrarily adjusted.
Furthermore, coupling large numbers of cavities resonantly has already been demonstrated on chips \cite{notomi2008large,hafezi2013imaging}, which indicates the potential for scaling up our system into large quantum circuits.

\noindent\textbf{Methods}\\
\textbf{Two-photon $S-$matrix and quantum correlations}
The few-photon scattering amplitude via a waveguide-coupled $\chi^{(2)}$ cavity can be calculated using a quantum field theoretical approach \cite{wang2022few} and, for the case of two incoming photons in the fundamental mode $a$, the result is given by (SI)
\begin{widetext}
\begin{equation}
\begin{aligned}\label{M-Smatrix}
S_{\nu_1\nu_2;\omega_1\omega_2}=& t_{\omega_1}t_{\omega_2}[\delta(\nu_1-\omega_1)\delta(\nu_2-\omega_2)+\delta(\nu_1-\omega_2)\delta(\nu_2-\omega_1)]\\
&-\kappa_{ae}^2\frac{2ig^2}{\pi}\frac{\omega_1+\omega_2-2\alpha_a}{(\omega_1-\alpha_a)(\omega_2-\alpha_a)(\nu_1-\alpha_a)(\nu_2-\alpha_a)(\omega_1+\omega_2-\lambda_1)(\omega_1+\omega_2-\lambda_2)}\delta(\nu_1+\nu_2-\omega_1-\omega_2),
\end{aligned}
\end{equation}
\end{widetext}
where $\alpha_{a(b)}=\omega_{a(b)}-i\frac{\kappa_{a(b)}}{2}$ and $\lambda_{1,2}=\frac{1}{2}(2\alpha_a+\alpha_b)\pm\frac{1}{2}\sqrt{(2\alpha_a-\alpha_b)^2+8g^2}$ are the eigenfrequencies of the closed Hilbert space of $\{ \ket{2_a0_b},\ket{0_a1_b}\}$ in the presence of interaction $\hat H_{\textrm{int}}= \hbar g(\hat a^{\dagger 2}\hat b+\hat a^2\hat b^\dagger)$.
For monochromatic input photons with frequency $\omega$, the temporal two-photon wavefunction of the transported photons is related to the $S-$matrix by 
\begin{equation}
\begin{aligned}
\psi_2(t,t+\tau)&=\int d\nu_1d\nu_2 S_{\nu_1\nu_2;\omega\omega} e^{-i\nu_1t-i\nu_2(t+\tau)}\\
&=e^{-i\omega(2t+\tau)}[t_\omega^2+T(\omega,\tau)],
\end{aligned}
\end{equation}
where $t_\omega$ is the single-photon transmission coefficient and 
\begin{equation}\label{T}
T(\omega,\tau)=-\frac{2g^2\kappa_{1e}^2}{(2\omega-\lambda_1)(2\omega-\lambda_2)(\omega-\alpha_a)^2}e^{-i|\tau|(\alpha_a-\omega)}.
\end{equation} 
The second-order correlation function thus is given by $g^{(2)}(\tau)\equiv\frac{|\psi_2(t,t+\tau)|^2}{|\psi_1(t)|^4}$, where $\psi_1(t)=e^{-i\omega t}t_\omega$ is the single-photon wavefunction. For the split-resonance cavity with waveguide feedback, the single-photon transmission near the local transmission minimum can be modeled as $t_\omega=r-\frac{i\kappa_{a,e}}{\omega-\omega_{\rm{min}}+i\kappa_a/2}$, where $\omega_{\rm{min}}$ is the frequency of the local minimum and is different from the bare cavity frequency $\omega_a$ due to the waveguide feedback. 
 \\
\textbf{Device fabrication}
The devices are fabricated from the 115 nm thick In$_{0.48}$Ga$_{0.52}$P thin film grown on GaAs substrate (0 degree off-cut toward [110]) by metal-organic chemical vapor deposition. The device pattern is defined using electron beam lithography and negative tone resist hydrogen silsesquioxane. The device pattern is transferred to InGaP layer via inductively coupled plasma reactive-ion etch (ICP-RIE) using a mixture of Cl$_2$/CH$_4$/Ar gas. A layer of 35 nm thick aluminum oxide is deposited on the chip via atomic layer deposition. This layer serves as the mechanical support for the suspended device. A second electron beam lithography and subsequent ICP-RIE using BCl$_3$ gas are applied to pattern etch-through holes in the aluminum oxide layer for the undercut of the InGaP device. A third electron beam lithography followed by electron-beam evaporation of 5 nm thick chromium and 20 nm thick gold is performed to define the high-resistance wires next to the 1550-nm light waveguide as the resistive phase shifter. A fourth electron beam lithography followed by electron-beam evaporation of 10 nm thick chromium and 150 nm thick gold is used to define the low-resistance large metal pads that connect the resistive heater to the wire bonding pad.  Finally, the InGaP device is released from the GaAs substrate using citric acid-based selective etching.\\
\textbf{InGaP photonic integrated circuit}
The design of the phase-matched microring resonator follows Ref. \cite{zhao2022ingap}. The nominal ring radius is 5 $\mu$m. The microring resonator is coupled to two bus waveguides for transmitting the 1550-nm and 775-nm wavelength light, respectively. The two bus waveguides are joined at a wavelength-division multiplexer (WDM), which is then connected to a tapered fiber-optic coupler which couples both 1550-nm TE polarized and 775-nm TM polarized light from a tapered fiber into the photonic circuit with an efficiency of about $60\%$ and $20\%$, respectively. At the end of the 1550-nm light waveguide, metal wires are fabricated on both sides of the waveguide in order to tune the phase of the waveguide via resistive heating and thermo-optic effect.  A group of 50 microrings with width increment at 1 nm step are fabricated in one device which effectively enhances the probability to realize the phase- and frequency-matching condition of a device and simplifies the measurement. \\
\textbf{Photon correlation measurement}   
The device chip is wire-bonded to a printed circuit board for voltage tuning of the waveguide phase and is mounted in the mixing chamber of a dilution refrigerator (DR) operated at 4 K. A bare optical fiber is fed through the DR for transmitting light. The tapered fiber tip is glued to the on-chip waveguide coupler for low-loss, mechanically-stable fiber-optic coupling that is immune to the DR vibration.
Light from a tunable external cavity diode laser (New Focus TLB-6728) is filtered by an optical grating filter (JDS TB9223, 3 dB bandwidth 0.55 nm, 20 dB bandwidth 1.5 nm) and a fiber Febry-Perot (FP) filter (LUNA, 50 GHz free spectral range, 400 finesse) to eliminate the amplified spontaneous emission and side-mode emission of the laser. One resonance of the FP filter is locked to the laser frequency via a dither feedback controller. The filtered light is then passed through a fiber polarizer (50 dB extinction) to eliminate unpolarized light and a 1550-nm/775-nm WDM before coupling into the device. The 775-nm wavelength second-harmonic light from the device is detected by an avalanche photodetector (Thorlabs APD440A). The 1550-nm wavelength light is further purified by a polarizer (40 dB extinction) and its correlations is analyzed using the Hanbury Brown-Twiss setup which consists of a 50:50 beamsplitter, two SNSPDs (Quantum Opus), and a TCSPC (Time Tagger Ultra).

\vspace{2mm}
\noindent\textbf{Data availability}\\
Data supporting the findings of this study are available within the article and its Supplementary Information, or from the corresponding author upon reasonable request.

\noindent\textbf{Acknowledgements}\\
We acknowledge the help from Siyuan Wang and Hao Tong during the device fabrication and measurement. We also thank Elizabeth Goldschmidt and Paul Kwiat for valuable discussions. This work was supported by US National Science Foundation (Grant No. DMS-1839177, ECCS-2223192) and U.S. Department of Energy Office of Science National Quantum Information Science Research Centers. S.F. acknowledges the support of a MURI project from the U. S. Air Force Office of Scientific Research (Grant No. FA9550-22-1-0339). 

\noindent\textbf{Author contributions}\\ 
Y.W., K.F., S.F. developed the theory. K.F. conceived the experiment. M.Z. fabricated the device. M.Z., K.F. performed the experiment and analyzed the result. K.F., S.F., M.Z. wrote the manuscript.

%\noindent\textbf{Additional information}\\
%Supplementary information is available in the online version of the paper.  Reprints and permissions are available at www.nature.com/reprints.  The authors declare no competing financial interests.  Correspondence and requests for materials should be sent to KF (kfang3@illinois.edu). \\
%
%\noindent\textbf{Competing financial interests}\\
%The authors declare no competing financial interests.
 
\end{document}

\title{Supplementary Information for: ``Quantum correlated photons via a passive nonlinear microcavity''}

\author{Mengdi Zhao} 
%\thanks{These authors contributed equally to this work.}
\affiliation{Holonyak Micro and Nanotechnology Laboratory, University of Illinois at Urbana-Champaign, Urbana, IL 61801 USA}
\affiliation{Department of Physics, University of Illinois at Urbana-Champaign, Urbana, IL 61801 USA}
\affiliation{Illinois Quantum Information Science and Technology Center, University of Illinois at Urbana-Champaign, Urbana, IL 61801 USA}
\author{Yunkai Wang} 
%\thanks{These authors contributed equally to this work.}
\affiliation{Holonyak Micro and Nanotechnology Laboratory, University of Illinois at Urbana-Champaign, Urbana, IL 61801 USA}
\affiliation{Department of Physics, University of Illinois at Urbana-Champaign, Urbana, IL 61801 USA}
\affiliation{Illinois Quantum Information Science and Technology Center, University of Illinois at Urbana-Champaign, Urbana, IL 61801 USA}
\author{Shanhui Fan} 
%\thanks{These authors contributed equally to this work.}
\affiliation{Department of Electrical Engineering, Ginzton Laboratory, Stanford University, Stanford, CA 94305, USA}
\author{Kejie Fang} 
\email{kfang3@illinois.edu}
%\homepage{https://fang.ece.illinois.edu}
\affiliation{Holonyak Micro and Nanotechnology Laboratory, University of Illinois at Urbana-Champaign, Urbana, IL 61801 USA}
\affiliation{Illinois Quantum Information Science and Technology Center, University of Illinois at Urbana-Champaign, Urbana, IL 61801 USA}
\affiliation{Department of Electrical and Computer Engineering, University of Illinois at Urbana-Champaign, Urbana, IL 61801 USA}

\begin{abstract}

\end{abstract}
%\pacs{}

\maketitle

\tableofcontents

\newpage

\section{Theory of two-photon transport}\label{App:theory}
\subsection{Computation of scattering amplitudes}\label{app:sub:Smatrix}
The $S-$matrix of few-photon transport via a waveguide-coupled nonlinear optical cavity can be systematically computed using quantum field theoretical methods \cite{xu2015input, wang2022few}, including the Feynman diagram approach suitable for the weak-coupling regime, i.e., $g/\kappa<1$. The purpose of this Section is to outline the key steps in calculation of two-photon scattering amplitude using the Feynman diagram approach. More details can be found in Ref. \cite{wang2022few}.

For a $\chi^{(2)}$ nonlinear cavity with an interaction Hamilton $H_{\textrm{int}}=\hbar g(a^{\dagger 2}b+a^2b^\dagger)$ and coupled with a one-port waveguide with an input-output relation $a_{\textrm{out}}(t)=a_{\textrm{in}}(t)-i\sqrt{\kappa_{ae}}a(t)$, the time-domain $S-$matrix of single-photon and two-photon transport is given by
\begin{equation}\label{S2t}
\begin{aligned}
S_{t;t'}&\equiv\bra{0}a_{\textrm{out}}(t)a_{\textrm{in}}^\dagger(t')\ket{0}\\
&=\delta(t-t')-\kappa_{ae}G(t;t')
\end{aligned}
\end{equation}
and 
%%\begin{widetext}
\begin{equation}\label{S4t}
\begin{aligned}
S_{t_1t_2;t'_1t'_2}&\equiv\bra{0}a_{\textrm{out}}(t_1)a_{\textrm{out}}(t_2)a_{\textrm{in}}^\dagger(t_1')a_{\textrm{in}}^\dagger(t_2')\ket{0}\\
&=\delta(t_1-t_1')\delta(t_2-t_2')+\delta(t_1-t_2')\delta(t_2-t_1')\\
&\quad-\kappa_{ae}G(t_1;t_1')\delta(t_2-t_2')-\kappa_{ae}G(t_2;t_2')\delta(t_1-t_1')-\kappa_{ae}G(t_1;t_2')\delta(t_2-t_1')-\kappa_{ae}G(t_2;t_1')\delta(t_1-t_2')\\
&\quad+\kappa_{ae}^2G(t_1,t_2;t_1',t_2'),
\end{aligned}
\end{equation}
%%\end{widetext}
respectively, where $G(t;t')\equiv\bra{0}\hat{T}[a(t)a^\dagger(t')]\ket{0}$ and $G(t_1,t_2;t_1',t_2')\equiv\bra{0}\hat{T}[a(t_1)a(t_2)a^\dagger(t_1')a^\dagger(t_2')]\ket{0}$ are the two-point and four-point Green's function of the cavity mode $a$, respectively, and $\hat{T}$ is the time-ordering operator.

The frequency-domain $S$-matrix characterizing the scattering amplitude between monochromatic output and input photons is obtained by the Fourier transform of the time-domain $S$-matrix, i.e.,
\begin{equation}\begin{aligned}\label{S2p}
S_{\nu_1;\omega_1}&\equiv \mathscr{F}\left[  S_{t;t'} \right] \\&= \int \frac{d t}{\sqrt{2 \pi}} e^{i \nu_{1} t} \int \frac{d t'}{\sqrt{2 \pi}} e^{i \omega_{1} t'} S_{t;t'} \\
& =\delta(\nu_1-\omega_1)-\kappa_{ae}G(\nu_1;\omega_1),
\end{aligned}\end{equation}
%and
%\begin{widetext}
\begin{equation}\label{S4p}
\begin{aligned}
S_{\nu_1\nu_2;\omega_1\omega_2}&\equiv \mathscr{F}\left[  S_{t_1t_2;t'_1t'_2} \right] \\
&=\delta(\nu_1-\omega_1)\delta(\nu_2-\omega_2)+\delta(\nu_1-\omega_2)\delta(\nu_2-\omega_1)\\
&\quad-\kappa_{ae}G(\nu_1;\omega_1)\delta(\nu_2-\omega_2)-\kappa_{ae}G(\nu_2;\omega_2)\delta(\nu_1-\omega_1)-\kappa_{ae}G(\nu_1;\omega_2)\delta(\nu_2-\omega_1)-\kappa_{ae}G(\nu_2;\omega_1)\delta(\nu_1-\omega_2)\\
&\quad+\kappa_{ae}^2G(\nu_1,\nu_2;\omega_1,\omega_2),
\end{aligned}
\end{equation}
%\end{widetext}
where $G(\nu_1;\omega_1)$ and $G(\nu_1,\nu_2;\omega_1,\omega_2)$ are the Fourier transform of the corresponding time-domain Green's functions.

\begin{figure}[!htb]
\begin{center}
\includegraphics[width=0.5\columnwidth]{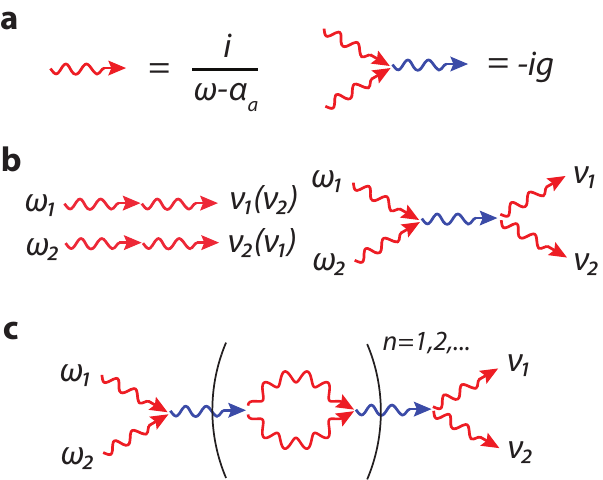}
\caption{\textbf{a}. Propagator and vertex of $\chi^{(2)}$ interaction.  \textbf{b} Leading-order Feynman diagrams for two-photon scattering. \textbf{c} Higher-order Feynman diagrams involved in two-photon scattering. The part in the bracket is repeated for $n$ times, $n\geq 1$. }
\label{fig:Feynman}
\end{center}
\end{figure}

The frequency-domain Green's function can be computed using Feynman diagrams. In essence, the $n-$point Green's function is calculated using all connected Feynman diagrams with $n$ external points, which are constructed from basic elements including the ``propagator'' and ``vertex''. The propagator is the two-point free-particle Green's function and the vertex corresponds to the bare interaction. The Feynman diagram rules for the $\chi^{(2)}$ cavity are given below:
\begin{itemize}
\item Propagator: $\frac{i}{\omega-\alpha_{a(b)}}$, where $\alpha_{a(b)}\equiv \omega_{a(b)}-i\frac{\kappa_{a(b)}}{2}$,
\item Vertex: $-ig$,
\item Impose energy conservation at each vertex: $\frac{1}{\sqrt{2\pi}}\delta(\sum \omega-\sum \nu)$,
\item Integrate undetermined momentum: $\int d\omega\int d\nu$. 
\item Multiply the symmetry factor: $m!$ for $m$ propagators of the same mode connecting two vertices or connecting a vertex with $m$ external points. 
\end{itemize}
The Feynman diagrams corresponding to the propagator and vertex are shown in Fig. \ref{fig:Feynman}a. 

The two-point Green's function (to the leading order) thus is
\begin{equation}\label{green_1}
G(\nu_1;\omega_1)=\frac{i}{\omega_1-\alpha_a}\delta(\nu_1-\omega_1).
\end{equation}
As a result, according to Eq. \ref{S2p}, we have
\begin{equation}\label{S11d}
S_{\nu_1;\omega_1}=\left[ 1-\frac{i\kappa_{ae}}{\omega_1-\alpha_a}\right]\delta(\nu_1-\omega_1)\equiv t_{\omega_1}\delta(\nu_1-\omega_1),
\end{equation}
where $t_{\omega_1}$ is the single-photon transmission coefficient. 

Fig. \ref{fig:Feynman}b shows the leading-order Feynman diagrams that contribute to the four-point Green's function. The first diagram is the uncorrelated transport of the two photons. The second diagram, representing the second-harmonic photon-mediated scattering, is calculated following the Feynman diagram rule as:
%\begin{widetext}
\begin{equation}
\begin{aligned}
&(2!)^2\frac{i}{\omega_1-\alpha_a}\frac{i}{\omega_2-\alpha_a}\frac{-ig}{\sqrt{2\pi}}\int dq\delta(q-\omega_1-\omega_2)\frac{i}{q-\alpha_b}\frac{-ig}{\sqrt{2\pi}}\delta(q-\nu_1-\nu_2)\frac{i}{\nu_1-\alpha_a}\frac{i}{\nu_2-\alpha_a}\\
=&-\frac{2ig^2}{\pi}\frac{1}{(\omega_1-\alpha_a)(\omega_2-\alpha_a)(\nu_1-\alpha_a)(\nu_2-\alpha_a)(\omega_1+\omega_2-\alpha_b)} \delta(\nu_1+\nu_2-\omega_1-\omega_2).
\end{aligned}
\end{equation}
Thus, the four-point Green's function and two-photon scattering matrix are found to be
\begin{equation}
\begin{aligned}\label{FG22}
G(\nu_1,\nu_2;\omega_1,\omega_2)=&-\frac{1}{(\omega_1-\alpha_a)(\omega_2-\alpha_a)}[\delta(\nu_1-\omega_1)\delta(\nu_2-\omega_2)+\delta(\nu_1-\omega_2)\delta(\nu_2-\omega_1)]\\&-\frac{2ig^2}{\pi}\frac{1}{(\omega_1-\alpha_a)(\omega_2-\alpha_a)(\nu_1-\alpha_a)(\nu_2-\alpha_a)(\omega_1+\omega_2-\alpha_b)} \delta(\nu_1+\nu_2-\omega_1-\omega_2)
\end{aligned}
\end{equation}
and
\begin{equation}\label{S22d}
\begin{aligned}
S_{\nu_1\nu_2;\omega_1\omega_2}=& t_{\omega_1}t_{\omega_2}[\delta(\nu_1-\omega_1)\delta(\nu_2-\omega_2)+\delta(\nu_1-\omega_2)\delta(\nu_2-\omega_1)]\\
&-\frac{2ig^2}{\pi}\frac{\kappa_{ae}^2}{(\omega_1-\alpha_a)(\omega_2-\alpha_a)(\nu_1-\alpha_a)(\nu_2-\alpha_a)(\omega_1+\omega_2-\alpha_b)} \delta(\nu_1+\nu_2-\omega_1-\omega_2),
\end{aligned}
\end{equation}
%\end{widetext}
respectively.

The exact expression of the $S-$matrix can be obtained by including all higher-order Feynman diagrams as shown in Fig. \ref{fig:Feynman}c, which yields
%\begin{widetext}
\begin{equation}
\begin{aligned}\label{S22exact}
S_{\nu_1\nu_2;\omega_1\omega_2}=& t_{\omega_1}t_{\omega_2}[\delta(\nu_1-\omega_1)\delta(\nu_2-\omega_2)+\delta(\nu_1-\omega_2)\delta(\nu_2-\omega_1)]\\
&-\kappa_{ae}^2\frac{2ig^2}{\pi}\frac{\omega_1+\omega_2-2\alpha_a}{(\omega_1-\alpha_a)(\omega_2-\alpha_a)(\nu_1-\alpha_a)(\nu_2-\alpha_a)(\omega_1+\omega_2-\lambda_1)(\omega_1+\omega_2-\lambda_2)}\delta(\nu_1+\nu_2-\omega_1-\omega_2),
\end{aligned}
\end{equation}
%\end{widetext}
where $\lambda_{1,2}=\frac{1}{2}(2\alpha_a+\alpha_b)\pm\frac{1}{2}\sqrt{(2\alpha_a-\alpha_b)^2+8g^2}$ are the eigenfrequencies of the closed Hilbert space of $\{ \ket{2_a0_b},\ket{0_a1_b}\}$ in the presence of interaction $H_{\mathrm{int}}=\hbar g(a^{\dagger 2}b+a^2b^\dagger)$. It can be straightforwardly verified that Eq. \ref{S22d} contains the leading-order terms up to $O(g^2/\kappa^2)$ of Eq. \ref{S22exact}. The exact form of $S-$matrix of Eq. \ref{S22exact} can be applied to both weak and strong coupling.

The result of the $S$-matrix here applies to generic waveguide-coupled cavity systems, where $t_\omega$ may not take the form of Eq. \ref{S11d} but depends on the specific construction of the photonic circuit.

\subsection{Second-order correlation function}
Here we calculate the second-order correlation function of the transported photons via the $\chi^{(2)}$ nonlinear circuit for a weak coherent state input
\begin{equation}\label{weakalpha}
\ket{\psi_{\textrm{in}}}=\ket{0}+\alpha\ket{1_t}+\frac{\alpha^2}{\sqrt{2}}\ket{1_t1_t}+O(\alpha^3),
\end{equation}
where $\alpha$ is the amplitude of the coherent state and $\ket{1_t}=\ket{1}e^{-i\omega t}$ represents a monochromatic single-photon state with frequency $\omega$ in the time domain. The second-order correlation function of the transported photons is given by
\begin{equation}
\begin{aligned}\label{g2smdef}
g^{(2)}(\tau)=\frac{\bra{\psi_{\textrm{in}}}a_{\textrm{out}}^\dagger(t)a_{\textrm{out}}^\dagger(t+\tau)a_{\textrm{out}}(t+\tau)a_{\textrm{out}}(t)\ket{\psi_{\textrm{in}}}}{|\bra{\psi_{\textrm{in}}}a_{\textrm{out}}^\dagger(t)a_{\textrm{out}}(t)\ket{\psi_{\textrm{in}}}|^2}.
\end{aligned}
\end{equation}
Using Eq. \ref{weakalpha} and keeping the leading order terms in $\ket{\psi_{\textrm{in}}}$, we have
\begin{equation}\begin{aligned}
&\bra{\psi_{\textrm{in}}}a_{\textrm{out}}^\dagger(t)a_{\textrm{out}}(t)\ket{\psi_{\textrm{in}}}\\
=&|\alpha|^2\bra{1_t} a_{\textrm{out}}^\dagger(t)a_{\textrm{out}}(t)\ket{1_t}\\
=&|\alpha|^2\bra{1_t} a_{\textrm{out}}^\dagger(t)\ket{0}\bra{0}a_{\textrm{out}}(t)\ket{1_t}\\
=&|\alpha|^2\int dp e^{i\nu t}\bra{1_\omega} a_{\textrm{out}}^\dagger(\nu)\ket{0} \int d\tilde \nu e^{-i\tilde \nu t}\bra{0}a_{\textrm{out}}(\tilde \nu)\ket{1_\omega}\\
=&|\alpha|^2\int d\nu e^{i\nu t}S_{\nu;\omega}^* \int d\tilde \nu e^{-i\tilde \nu t}S_{\tilde \nu;\omega} \\
=&|\alpha|^2|t_\omega|^2
\end{aligned}\end{equation}
and
\begin{equation}\begin{aligned}\label{g2num}
&\bra{\psi_{\textrm{in}}}a_{\textrm{out}}^\dagger(t)a_{\textrm{out}}^\dagger(t+\tau)a_{\textrm{out}}(t+\tau)a_{\textrm{out}}(t)\ket{\psi_{\textrm{in}}}\\
=&\frac{|\alpha|^4}{2}\bra{1_t1_t}a_{\textrm{out}}^\dagger(t)a_{\textrm{out}}^\dagger(t+\tau)a_{\textrm{out}}(t+\tau)a_{\textrm{out},1}(t)\ket{1_t1_t}\\
=&\frac{|\alpha|^4}{2}\int d\nu_1d\nu_2 \bra{1_\omega1_\omega}a_{\textrm{out}}^\dagger(\nu_1)a_{\textrm{out}}^\dagger(\nu_2)\ket{0}  e^{i\nu_1t+i\nu_2(t+\tau)}\\
&\quad\times\int d\tilde \nu_1d\tilde \nu_2\bra{0}a_{\textrm{out}}(\tilde \nu_2)a_{\textrm{out}}(\tilde \nu_1)\ket{1_\omega1_\omega}e^{-i\tilde \nu_1t-i\tilde \nu_2(t+\tau)}\\
=& \frac{|\alpha|^4}{4}\int d\nu_1d\nu_2 S_{\nu_1\nu_2;\omega\omega}^* e^{i\nu_1t+i\nu_2(t+\tau)}\\
&\quad \times \int d\tilde \nu_1d\tilde \nu_2 S_{\tilde \nu_1\tilde \nu_2;\omega\omega}e^{-i\tilde \nu_1t-i\tilde \nu_2(t+\tau)} \\
=&|\alpha|^4|t_\omega^2+T(\omega,\tau)|^2,
\end{aligned}\end{equation}
where the $S-$matrices are given by Eqs. \ref{S11d} and \ref{S22exact} and
\begin{equation}\begin{aligned}\label{Tktaud}
T(\omega,\tau)=-\frac{2g^2\kappa_{1e}^2}{(2\omega-\lambda_1)(2\omega-\lambda_2)(\omega-\alpha_a)^2}e^{-i|\tau|(\alpha_a-\omega)}.
\end{aligned}\end{equation}
As a result, we have
\begin{equation}\begin{aligned}\label{g2result}
g^{(2)}(\tau)=\frac{|t_\omega^2+T(\omega,\tau)|^2}{|t_\omega^2|^2}.
\end{aligned}\end{equation}

%\color{NavyBlue}

\begin{figure*}[!htb]
\begin{center}
\includegraphics[width=\columnwidth]{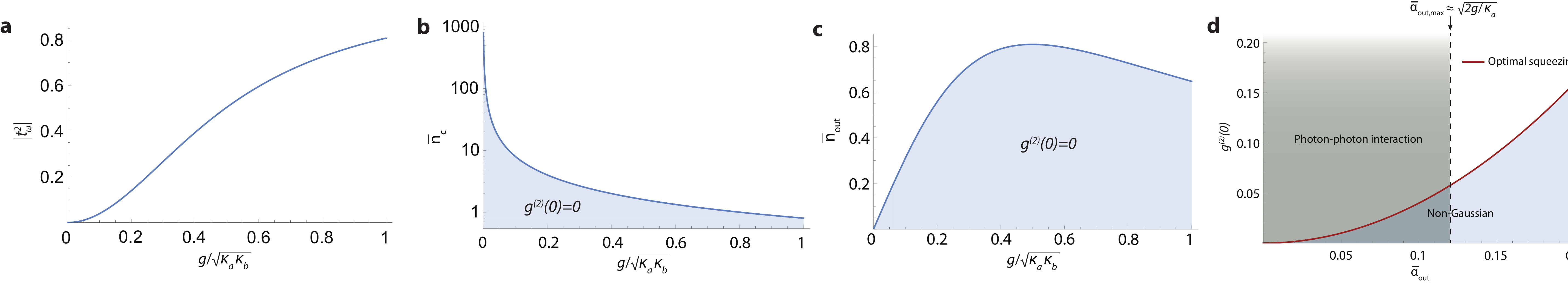}
\caption{\textbf{a}. Relation between $t_\omega^2$ and $g/\sqrt{\kappa_a\kappa_b}$ for realizing $g^{(2)}(0)=0$. \textbf{b}. Range of mean cavity photon number $\bar n_c$ allowed for $g^{(2)}(0)=0$. \textbf{c}. Range of output  photon number $\bar n_{c,\rm out}$ allowed for $g^{(2)}(0)=0$. These calculations use $\kappa_a/2\pi=1.6$ GHz and $\kappa_b/2\pi=4$ GHz. \textbf{d} Achievable $g^{(2)}(0)\ll 1$ via the photon-photon interaction in the $\chi^{(2)}$ cavity (dark grey area). $g^{(2)}(0)=0$ can be achieved up to an approximate maximum output field amplitude $\bar\alpha_{\rm out,max}\approx\sqrt{\frac{2g}{\kappa_a}}$. Red curve corresponding to the optimal squeezing given by $g^{(2)}(0)= 4|\bar \alpha_{\rm out}|^2$. The light blue area represents non-Gaussian state regime.}
\label{fig:g2cavityphoton}
\end{center}
\end{figure*}

It is worth discussing when $g^{(2)}(0)=0$ can be achieved. 
For general $g$, Fig. \ref{fig:g2cavityphoton}a plots the relation between $t_\omega^2$ and $g/\sqrt{\kappa_a\kappa_b}$ to achieve $g^{(2)}(0)=0$ using Eq. \ref{g2result}, where we have used practical photon dissipation rate $\kappa_a/2\pi=1.6$ GHz and $\kappa_b/2\pi=4$ GHz as in our device. Our theory is valid given the mean second-harmonic (SH) cavity photon number satisfies $\bar n_{c,\rm SH}\lesssim1$.  
Thus, $\bar n_{c,\rm SH}=1$ imposes an upper bound of the input power and fundamental-mode cavity photon number for observation of the quantum correlations. For the latter, the bound is $\bar n_{c, \rm max}=\frac{\kappa_b}{2g}$. Fig. \ref{fig:g2cavityphoton}b show the viable range of $\bar n_c$ for observing $g^{(2)}(0)=0$. It is perhaps more illustrative to show the output photon number, defined as $\bar n_{\rm out}\equiv t_\omega^2\times 2P_{\rm in}/\hbar\omega_a/\kappa_a=\bar n_c t_\omega^2$, where we have assumed the output photons are fed into a cavity of linewidth $\kappa_a$. Fig. \ref{fig:g2cavityphoton}c shows the viable range of $\bar n_{\rm out}$ that could be associated with $g^{(2)}(0)=0$, using the result of Fig. \ref{fig:g2cavityphoton}a and b. 

These results highlight the difference between our scheme and the so-called unconventional photon blockade (UPB) (see e.g. \cite{flayac2017unconventional} and recent experimental demonstrations based on quantum dots \cite{snijders2018observation} and nonlinear microwave resonators \cite{vaneph2018observation}). UPB explores weak nonlinearities to achieve non-classical correlations and even photon blockade. Usually multiple pumps are needed in UPB and the pumps need to be tuned in order to achieve $g^{(2)}(0)<1$. For UPB, the minimum achievable $g^{(2)}(0)$ is dependent on the photon number and usually $g^{(2)}(0)=0$ is only achieved for $\bar n\rightarrow 0$. In contrast, in our scheme, the minimum achievable $g^{(2)}(0)$ is independent of the input power or cavity photon number (up to the single-intermediate-photon bound).  

To be more specific, we compare with the UPB scheme based on a doubly-resonant $\chi^{(2)}$ cavity \cite{zhou2015unconventional}. There, both the fundamental and second-harmonic modes of a $\chi^{(2)}$ cavity are driven by external pumps. It is found, in order to achieve the optimal antibunching, the fundamental and second-harmonic cavity photon numbers need to satisfy $\bar n_{c,\rm SH}=\left(\frac{\kappa_a^2\bar n_c}{g\kappa_b}\right)^2$. This relation is different from the SHG given by $\bar n_{c,\rm SH}=4\left(\frac{g\bar n_c}{\kappa_b}\right)^2$, and as a result an external SH pump is needed for UPB. The same condition also means that for a fixed SH pump, optimal antibunching can only be achieved for a particular $\bar n_c$ given by $\frac{g\kappa_b}{\kappa_a^2}\sqrt{\bar n_{c,\rm SH}}$.

Later it was found that UPB can be explained by the ``optimal squeezing'' with a linearized Hamiltonian and thus the states generated by the UPB is Gaussian \cite{lemonde2014antibunching}. In contrast, in our scheme, the minimum achievable $g^{(2)}(0)$ is independent on the photon number up to a limit imposed by $\bar n_{c,\rm SH}\approx 1$, as shown in Fig. \ref{fig:g2cavityphoton}b and c. This means it is possible to create non-Gaussian states in our approach as illustrated in Fig. \ref{fig:g2cavityphoton}d. The red curve is the minimum achievable $g^{(2)}(0)$ via ``optimal squeezing'' given by $g^{(2)}(0)= 4|\bar \alpha_{\rm out}|^2$, where $|\bar \alpha_{\rm out}|\equiv \sqrt{\bar n_{\rm out}}$, below which is the non-Gaussian state regime \cite{lemonde2014antibunching}. In our approach, up to approximately $|\bar \alpha_{\rm out,max}|= \sqrt{\bar n_{c, \rm max}\times \frac{4g^2}{\kappa_a\kappa_b}}=\sqrt{\frac{2g}{\kappa_a}}$ (given $g\ll \kappa$), $g^{(2)}(0)=0$ can be achieved. Thus, non-Gaussian states can be created via the photon-photon interaction in the $\chi^{(2)}$ cavity.

\color{Black}
\subsection{Modeling of the effective one-port photonic circuit}\label{modeloneport}

\begin{figure*}[!htb]
\begin{center}
\includegraphics[width=0.6\columnwidth]{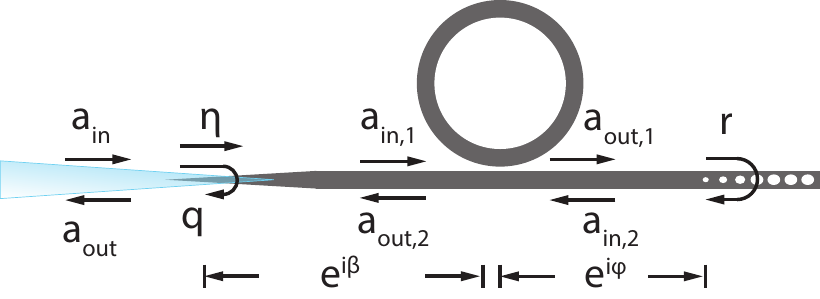}
\caption{Input-output relation of the effective one-port photonic circuit.}
\label{fig:oneport}
\end{center}
\end{figure*}

For an ideal ring resonator, the clockwise (CW) and counterclockwise (CCW) resonances are degenerate with frequency $\omega_0$. For the fabricated high-index InGaP ring resonator, the CW and CCW resonances couple due to the surface roughness induced back-scattering. As illustrated in Fig. \ref{fig:oneport}, the input-output equation for the CW and CCW resonances is given by 

%\begin{widetext}
\begin{equation}
\label{eq:eqofmotion}
\begin{aligned}
\frac{d}{dt}\left(\begin{array}{c} a_{\mr{cw}} \\  a_{\mr{ccw}} \end{array}\right) = -i\left(\begin{array}{cc} \omega_0-i\frac{\kappa_{\mr{cw}}}{2} & V\\ V & \omega_0-i\frac{\kappa_{\mr{ccw}}}{2} \end{array}\right)\left(\begin{array}{c} a_{\mr{cw}} \\  a_{\mr{ccw}} \end{array}\right) +\left(\begin{array}{cc} 0 & \sqrt{\kappa_{\mr{cw},e}}\\ \sqrt{\kappa_{\mr{ccw},e}} & 0 \end{array}\right)\left(\begin{array}{c} a_{\mr{in},1} \\  a_{\mr{in},2} \end{array}\right),
\end{aligned}
\end{equation}
\begin{equation}
\label{eq:inout}
\left(\begin{array}{c} a_{\mr{out},1} \\  a_{\mr{out},2} \end{array}\right)=\left(\begin{array}{c} a_{\mr{in},1} \\  a_{\mr{in},2} \end{array}\right)-\left(\begin{array}{cc} 0 & \sqrt{\kappa_{\mr{ccw},e}}\\ \sqrt{\kappa_{\mr{cw},e}} & 0 \end{array}\right)\left(\begin{array}{c} a_{\mr{cw}} \\  a_{\mr{ccw}} \end{array}\right),
\end{equation}
%\end{widetext}
where $\kappa_{\mr{cw(ccw)}}=\kappa_{\mr{cw(ccw)},e}+\kappa_{\mr{cw(ccw)},i}$ and $V$ is the coupling between the CW and CCW resonances.

The photonic crystal mirror with non-unity reflection acts like a beamsplitter, which relates $a_{\mr{out},1}$ and $a_{\mr{in},2}$ by
\begin{equation}
\label{eq:phc}
a_{\mr{in},2}=e^{i\varphi}(ir e^{i\varphi} a_{\mr{out},1}+t a_{n}),
\end{equation}
where $\varphi$ is the waveguide phase between the ring and the photonic crystal mirror, $r$ and $t$ are the real reflection and transmission coefficient of the photonic crystal mirror, and $a_{n}$ represents the noise operator.

By eliminating $a_{\mr{out},1}$ and $a_{\mr{in},2}$ from Eqs.\ref{eq:eqofmotion}-\ref{eq:phc}, we obtain
%\begin{widetext}
\begin{equation}
\label{eq:cwccw}
\begin{aligned}
& \frac{d}{dt}\left(\begin{array}{c} a_{\mr{cw}} \\  a_{\mr{ccw}} \end{array}\right)  =-i\left(\begin{array}{cc} \omega_0-i\frac{\kappa_{\mr{cw}}}{2} & V+\sqrt{\kappa_{\mr{cw},e}\kappa_{\mr{ccw},e}} r e^{i2\varphi}\\ V & \omega_0-i\frac{\kappa_{\mr{ccw}}}{2} \end{array}\right)\left(\begin{array}{c} a_{\mr{cw}} \\  a_{\mr{ccw}} \end{array}\right) +\left(\begin{array}{c} i\sqrt{\kappa_{\mr{cw},e}}re^{i2\varphi}\\  \sqrt{\kappa_{\mr{ccw},e}} \end{array}\right) a_{\mr{in},1} +\left(\begin{array}{c} \sqrt{\kappa_{\mr{cw},e}}te^{i\varphi}\\  0 \end{array}\right) a_{n},
\end{aligned}
\end{equation}
%\end{widetext}
\begin{equation}\label{eq:cwccwinout}
a_{\mr{out},2}=ire^{i2\varphi}a_{\mr{in},1}-\sqrt{\kappa_{\mr{cw},e}}a_{\mr{cw}}-i\sqrt{\kappa_{\mr{ccw},e}}r e^{i2\varphi}a_{\mr{ccw}}+t e^{i\varphi}a_{n}.
\end{equation}

From now on, we assume the CW and CCW resonances have the same dissipation rate and omit the label. 
To obtain the reflection coefficient of the device, we solve Eqs.\ref{eq:cwccw} and \ref{eq:cwccwinout} in the frequency domain by ignoring the noise operator, which yields
%\begin{widetext}
\begin{equation}
\label{eq:inout_rig}
a_{\mr{out},2}[\omega]=i\left[ r e^{i2\varphi}-\kappa_e \left(\begin{array}{cc} 1& ire^{i2\varphi} \end{array}\right) \left(\begin{array}{cc} \omega-\omega_0+i\frac{\kappa}{2} & -V-\kappa_{e}r e^{i2\varphi}\\ -V & \omega-\omega_0+i\frac{\kappa}{2} \end{array}\right)^{-1} \left(\begin{array}{c} ire^{i2\varphi}\\1 \end{array}\right) \right]a_{\mr{in},1}[\omega].
\end{equation}
%\end{widetext}
The reflection is zero (i.e., $a_{\mr{out},2}[\omega]=0$) when
\begin{equation}\label{R0c1}
\left(\cos2\varphi+\frac{r\Delta\kappa^2}{2V\kappa_e}\right)^2=1- \frac{r^2\Delta\kappa^2}{\kappa_e^2}
\end{equation}
and
\begin{equation}\label{R0c2}
\begin{aligned}
\omega&=\omega_0\pm\sqrt{V^2+\Delta\kappa^2+ V(\kappa_e/r)\cos2\varphi}\\
&=\omega_0\pm\sqrt{V^2+(\Delta\kappa/2)^2\pm V\sqrt{(\kappa_e/r)^2-\Delta\kappa^2}},
\end{aligned}
\end{equation}
where $\Delta\kappa=\kappa_i-\kappa_e$. It is obvious that the frequency corresponding to the zero reflection is not the bare cavity resonance frequency.

From now on, we consider the large splitting limit, i.e., $V\gg\kappa$, to simplify the two-mode modeling. By eliminating $a_{\mr{out},1}$ and $a_{\mr{in},2}$ from Eqs.\ref{eq:eqofmotion}-\ref{eq:phc},and diagonalizing Eq. \ref{eq:eqofmotion} , we obtain
%\begin{widetext}
\begin{equation}
\begin{aligned}
& \frac{d}{dt}\left(\begin{array}{c} a_{\mr{e}} \\  a_{\mr{o}} \end{array}\right)  =-i\left(\begin{array}{cc} \omega_{\mr{e}}-i\frac{\kappa_{\mr{e}} }{2} & \\  & \omega_{\mr{o}}-i\frac{\kappa_{\mr{o}} }{2} \end{array}\right)\left(\begin{array}{c} a_{\mr{e}} \\  a_{\mr{o}} \end{array}\right) +\frac{1}{\sqrt{2}}\left(\begin{array}{c} i\sqrt{\kappa_{e}}re^{i2\varphi}+\sqrt{\kappa_{e}} \\   i\sqrt{\kappa_{e}}re^{i2\varphi}-\sqrt{\kappa_{e}} \end{array}\right) a_{\mr{in},1} +\frac{1}{\sqrt{2}}\left(\begin{array}{c} \sqrt{\kappa_{e}}te^{i\varphi}\\  \sqrt{\kappa_{e}}te^{i\varphi} \end{array}\right) a_{n},
\end{aligned}
\end{equation}
\begin{equation}
a_{\mr{out},2}=ire^{i2\varphi}a_{\mr{in},1}-\frac{1}{\sqrt{2}}(i\sqrt{\kappa_{e}}re^{i2\varphi}+\sqrt{\kappa_{e}} )a_{\mr{e}}-\frac{1}{\sqrt{2}}(-i\sqrt{\kappa_{e}}re^{i2\varphi}+\sqrt{\kappa_{e}} )a_{\mr{o}}+t e^{i\varphi}a_{n},
\end{equation}
%\end{widetext}
where $a_{\mr{e(o)}}=\frac{1}{\sqrt{2}}(a_{\mr{cw}}\pm a_{\mr{ccw}})$, $\kappa_{\mr{e(o)}}=\kappa\mp r\kappa_e\sin2\varphi$, and 
\be\label{tunefreq}
\omega_{\mr{e(o)}}=\omega_0\pm (V+ \frac{1}{2}r\kappa_e\cos2\varphi).
\ee
As a result, in the large splitting limit, the hybridized even and odd modes, $a_{\mr{e}}$ and $a_{\mr{o}}$, become independently described by the input-output equations
\begin{equation}
\frac{da_{\mr{o(e)}}}{dt}=-i(\omega_{\mr{o(e)}}-i\frac{\kappa_{\mr{o(e)}}}{2})a_{\mr{o(e)}}-\frac{\sqrt{\kappa_e}(\pm 1- ire^{i2\varphi})}{\sqrt{2}}a_{\mr{in},1},
\end{equation}
\begin{equation}
a_{\mr{out},2}=ire^{i2\varphi}a_{\mr{in},1}-\frac{\sqrt{\kappa_e}(1\mp ire^{i2\varphi})}{\sqrt{2}}a_{\mr{o(e)}}.
\end{equation}
Replacing $\frac{\sqrt{\kappa_e}(1 \mp ire^{i2\varphi})}{\sqrt{2}}$ with $\sqrt{\kappa_{\mr{(o)e,e}}}e^{i(\theta+\pi/4)}$, $a_{\mr{o(e)}} e^{-i(\theta-\pi/4)}$ with $a_{\mr{o(e)}}$, and $a_{\mr{out,2}}e^{-2i(\theta-\pi/4)}$ with $a_{\mr{out}}$, the input-output relations are simplified to
\begin{equation}
\frac{da_{\mr{o(e)}}}{dt}=-i(\omega_{\mr{o(e)}}-i\frac{\kappa_{\mr{o(e)}}}{2})a_{\mr{o(e)}}-i\sqrt{\kappa_{\mr{o(e)},e}}a_{\mr{in},1},
\end{equation}
\begin{equation}
a_{\mr{out},2}=Ra_{\mr{in},1}-i\sqrt{\kappa_{\mr{o(e)},e}}a_{\mr{o(e)}},
\end{equation}
where 
\begin{equation}
R=-re^{i(2\varphi-2\theta)}
\end{equation}
and the effective external and intrinsic dissipation rates of $a_{\mr{o(e)}}$ are
\be\label{eke}
\kappa_{\mr{e(o)},e}=\kappa_e\left(\frac{1+r^2}{2}\mp r\sin2\varphi\right)
\ee
and
\be\label{R0V}
\kappa_{\mr{e(o)},i}=\kappa_i+\frac{t^2}{2}\kappa_e.
\ee
After Fourier transform, the input-output relation in the frequency domain is obtained
\begin{equation}\begin{aligned}\label{tnocoupler}
a_{\mr{out},2}[\omega]&=\Big[R-\frac{i\kappa_{\mr{o(e)},e}}{\omega-\omega_{\mr{o(e)}}+i\kappa_{\mr{o(e)}}/2}\Big] a_{\mr{in},1}[\omega]\\
&\equiv \tilde t_\omega a_{\mr{in},1}[\omega].
\end{aligned}\end{equation}

The zero-reflection condition at the large splitting limit is
\begin{equation}\label{R0V}
\sin2\varphi=\pm \frac{r\Delta\kappa}{\kappa_e}
\end{equation}
and
\begin{equation}
\begin{aligned}\label{mintf}
\omega &=\omega_0\pm (V+ \frac{1}{2}\frac{\kappa_e}{r}\cos2\varphi)\\&=\omega_0\pm (V\pm \frac{1}{2}\sqrt{(\kappa_e/r)^2-\Delta\kappa^2}), 
\end{aligned}
\end{equation}
which is consistent with Eqs. \ref{R0c1} and \ref{R0c2}. From Eq. \ref{R0V}, it is seen that the zero-reflection condition can be achieved only if $\kappa_i \leqslant (1+\frac{1}{r})\kappa_e$. Note Eq. \ref{R0V} is not the same as $\kappa_{\mr{e(o)},e}=\kappa_{\mr{e(o)},i}$ for $r<1$ because of the Fano interference. The difference between Eqs. \ref{mintf} and \ref{tunefreq} gives the local minimum transmission-cavity resonance detuning in the large-splitting limit.

Next, we derive the fiber-to-fiber transmission of the signal, taking into account of the direct reflection at the fiber-coupler region. The transmission relation of the fiber coupler can be modeled as
\begin{equation}
a_{\mr{out}}=iq a_{\mr{in}}+e^{i\beta}\eta a_{\mr{out},2},
\end{equation}
\begin{equation}
a_{\mr{in},1}=e^{i\beta}\eta a_{\mr{in}}.
\end{equation}
Therefore, 
\begin{equation}
\begin{aligned}
a_{\mr{out}}[\omega]&=\Big[\eta^2e^{2i\beta}\Big(R-\frac{i\kappa_{\mr{o(e)},e}}{\omega-\omega_{\mr{o(e)}}+i\kappa_{\mr{o(e)}}/2}\Big)+iq\Big] a_{\mr{in}}[\omega]\\&\equiv t_\omega a_{\mr{in}}[\omega].
\end{aligned}
\end{equation}

The second-order correlation function, following the derivation in Section \ref{app:sub:Smatrix}, is given by
\begin{equation}\begin{aligned}
g^{(2)}(\tau)&=\frac{|t_\omega^2+\eta^4e^{4i\beta}T(\omega,\tau)|^2}{|t_\omega^2|^2}\\
&=\frac{|(R+iq\eta^{-2}e^{-2i\beta})^2\bar t_\omega^2+T(\omega,\tau)|^2}{|(R+iq\eta^{-2}e^{-2i\beta})^2\bar t_\omega^2|^2},
\end{aligned}\end{equation}
where 
\be
\bar t_\omega=1-\frac{1}{R+iq\eta^{-2}e^{-2i\beta}}\frac{i\kappa_{\mr{o(e)},e}}{\omega-\omega_{\mr{o(e)}}+i\kappa_{\mr{o(e)}}/2}
\ee
is the transmission coefficient normalized to the off-resonance background, i.e., the one that is measured and quoted throughout this paper. $T(\omega,\tau)$ is calculated using Eq. \ref{Tktaud} (note $\omega\equiv k$) with the effective dissipation rate given by Eqs. \ref{eke} and \ref{R0V}. 

If $q\ll |R|\eta^{2}$, which is satisfied for our devices (see Table \ref{tab:deviceparameter}), 
\begin{equation}\begin{aligned}
g^{(2)}(\tau)&\approx\frac{|R^2\bar t_\omega^2+T(\omega,\tau)|^2}{|R^2\bar t_\omega^2|^2}\\
&\approx\frac{|\tilde t_\omega^2+T(\omega,\tau)|^2}{|\tilde t_\omega^2|^2},
\end{aligned}\end{equation}
where $\tilde t_\omega$ is given by Eq. \ref{tnocoupler}.

In general, for the split resonances, the local minimum of the transmission spectrum is detuned from the bare cavity frequency. Thus, when the signal is at the local minimum of the transmission, this detuning induces an oscillation of $T(\omega, \tau)$ (Eq. \ref{Tktaud}), which is the cause of the repulsive photon-photon interaction.

\subsection{Tunable second-order quantum correlations}
In the case that the resonance splitting is less than the resonance linewidth, as in the measured InGaP device, we can use the following model for the second-order correlation function of near-critically coupled light, 
\begin{equation}\label{fitg2}
g^{(2)}(\tau)=\frac{|t_\omega^2+T(\omega,\tau)|^2}{| t_\omega^2|^2},
\end{equation}
\be\label{fitt2}
t_\omega=r-\frac{i\kappa_{a,e}}{\omega-\omega_{\rm{min}}+i\kappa_a/2}
\ee
and 
\begin{equation}\label{fitT}
T(\omega,\tau)=-\frac{g^2\kappa_{ae}^2}{(2\omega-\alpha_b)(\omega-\alpha_a)^3}e^{-i|\tau|(\alpha_a-\omega)}.
\end{equation} 
where $\omega_{\rm{min}}$ is the frequency of the local minimum and is different from the bare cavity frequency $\omega_a$ due to the waveguide feedback. 

\begin{figure*}[!htb]
\begin{center}
\includegraphics[width=\columnwidth]{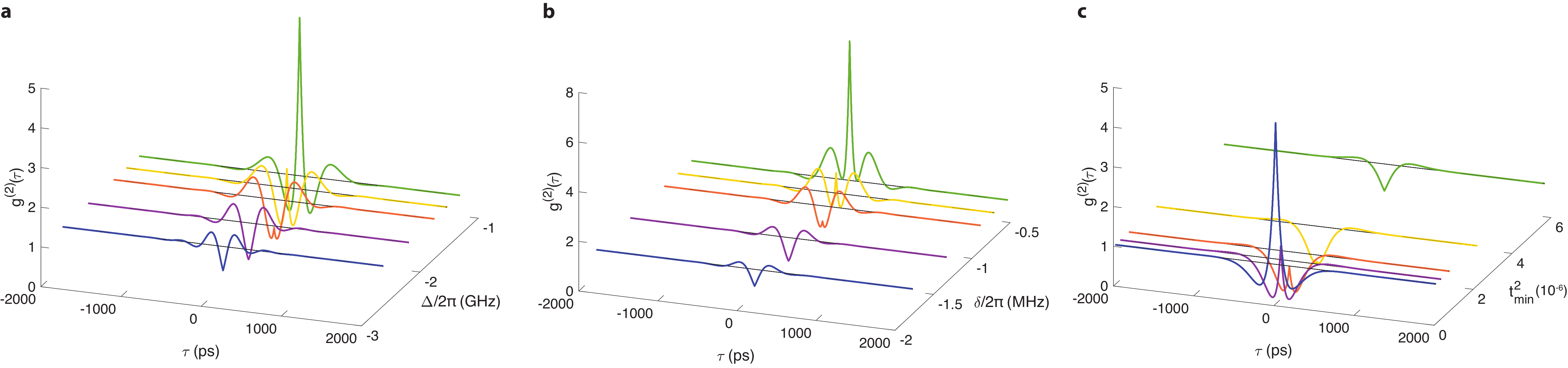}
\caption{ $g^{(2)}(\tau)$ for various tuning parameters. Black line indicates $g^{(2)}=1$. \textbf{a}. $t^2_{\textrm{min}}=2.1\times 10^{-7}$, $\delta/2\pi=0.88$ MHz. \textbf{b}. $t^2_{\textrm{min}}=1.4\times 10^{-7}$, $\Delta/2\pi=-1.9$ GHz. \textbf{c}. $\delta/2\pi=2.2$ MHz, $\Delta/2\pi=-0.68$ GHz. For these calculations, we assumed $\omega_b=2\omega_a$ and used the following parameters: $Q_{ai}=1.8\times 10^5$, $Q_{bi}=5\times 10^4$, $Q_{be}=2\times 10^6$, $g/2\pi=7.8$ MHz, and $r=1$.}
\label{fig:g2model}
\end{center}
\end{figure*}

The quantum correlation of transmitted photons can be controlled by the local transmission minimum, $t^2_{\rm{min}}$, the signal-local minimum detuning, $\delta\equiv\omega-\omega_{\rm{min}}$, and minimum transmission-cavity detuning, $\Delta\equiv\omega_a-\omega_{\rm{min}}$.
Fig. \ref{fig:g2model} plots the available $g^{(2)}(\tau)$ for various parameters of practical devices. 

\section{Device fabrication}\label{App:fab}

\begin{figure*}[!htb]
\begin{center}
\includegraphics[width=0.8\columnwidth]{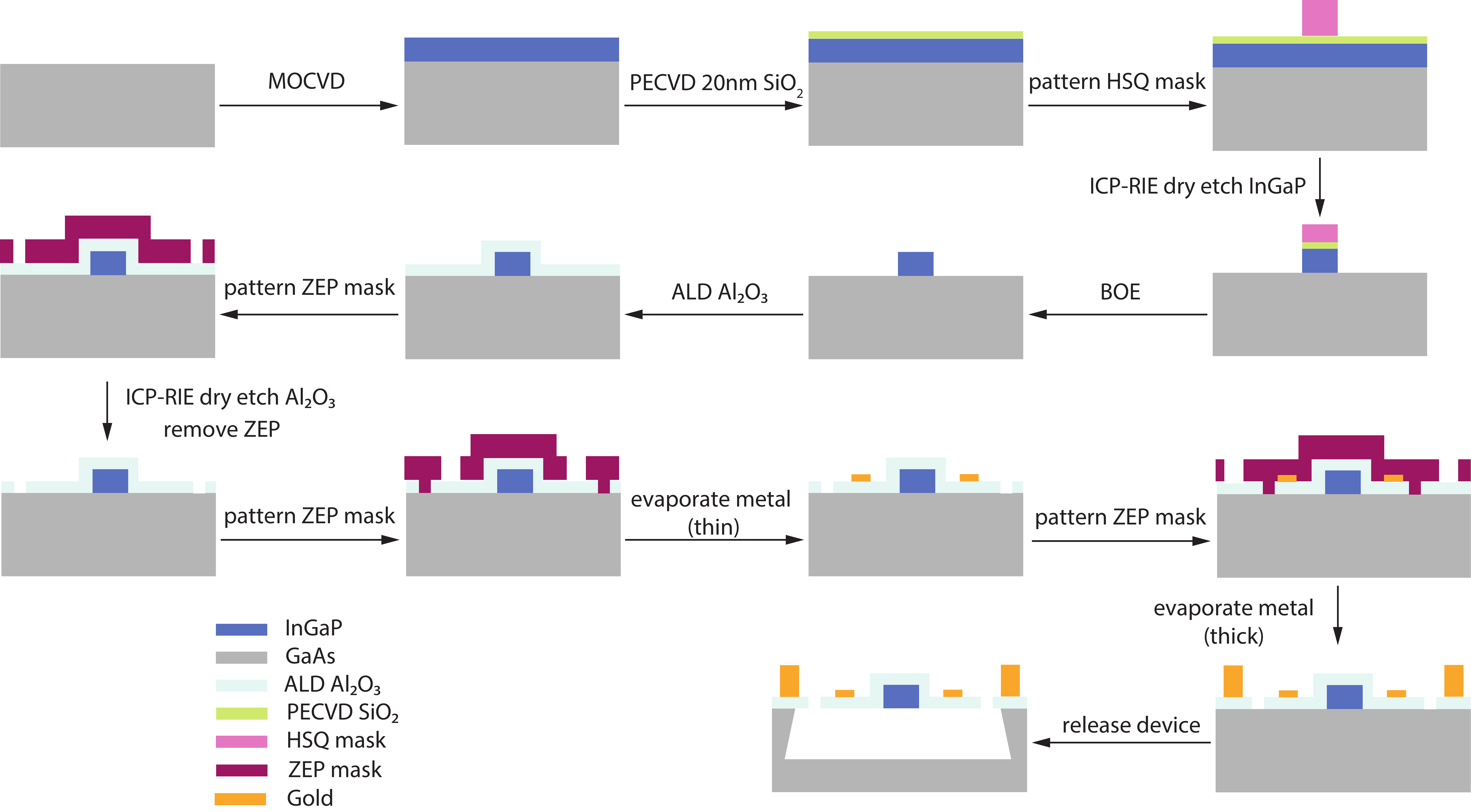}
\caption{Fabrication process of the InGaP photonic circuit. }
\label{fig:fab}
\end{center}
\end{figure*}

Fig. \ref{fig:fab} shows the flow chart of the fabrication process of the InGaP photonic circuit. The devices are fabricated from the 115 nm thick disordered In$_{0.48}$Ga$_{0.52}$P thin film grown on GaAs substrate by metal-organic chemical vapor deposition (T = 545 C, V/III = 48, precursors: trimethylindium, trimethylgallium and PH$_3$). The device pattern is defined using a 150 keV electron beam lithography tool and 150 nm thick negative tone resist hydrogen silsesquioxane (HSQ). A 20 nm thick layer of silicon dioxide is deposited on InGaP via plasma-enhanced chemical vapor deposition (PECVD) to promote the adhesion of HSQ. The device pattern is transferred to InGaP layer via inductively coupled plasma reactive-ion etch (ICP-RIE) using a mixture of Cl$_2$/CH$_4$/Ar gas with a selectivity of InGaP: HSQ: PECVD SiO$_2$ = 240: 90: 80. After a short buffered oxide etch to remove the residual oxide (both HSQ and PECVD oxide), a layer of 35 nm thick aluminum oxide is deposited on the chip via atomic layer deposition (ALD). This layer will serve as the mechanical support for the suspended device. A second electron beam lithography and subsequent ICP-RIE using BCl$_3$ gas are applied to pattern etch-through holes in the aluminum oxide layer for the undercut of the InGaP device. Next, a third electron beam lithography followed by electron-beam evaporation of 5 nm thick chromium and 20 nm thick gold is performed to define the high-resistance wires next to the 1550-nm light waveguide as the phase shifter. Then, a fourth electron beam lithography followed by electron-beam evaporation of 10 nm thick chromium and 150 nm thick gold is used to define the low-resistance large metal pads that connect several devices in parallel. In this way, the resistive heating is concentrated at the phase shifter without heating up the whole chip. Finally, the InGaP device is released from the GaAs substrate using citric acid-based selective etching. The suspended InGaP device is mechanically anchored to the aluminum oxide membrane.

\section{The InGaP photonic circuit: design and characterization}\label{App:PIC}

\begin{figure}[!htb]
\begin{center}
\includegraphics[width=0.5\columnwidth]{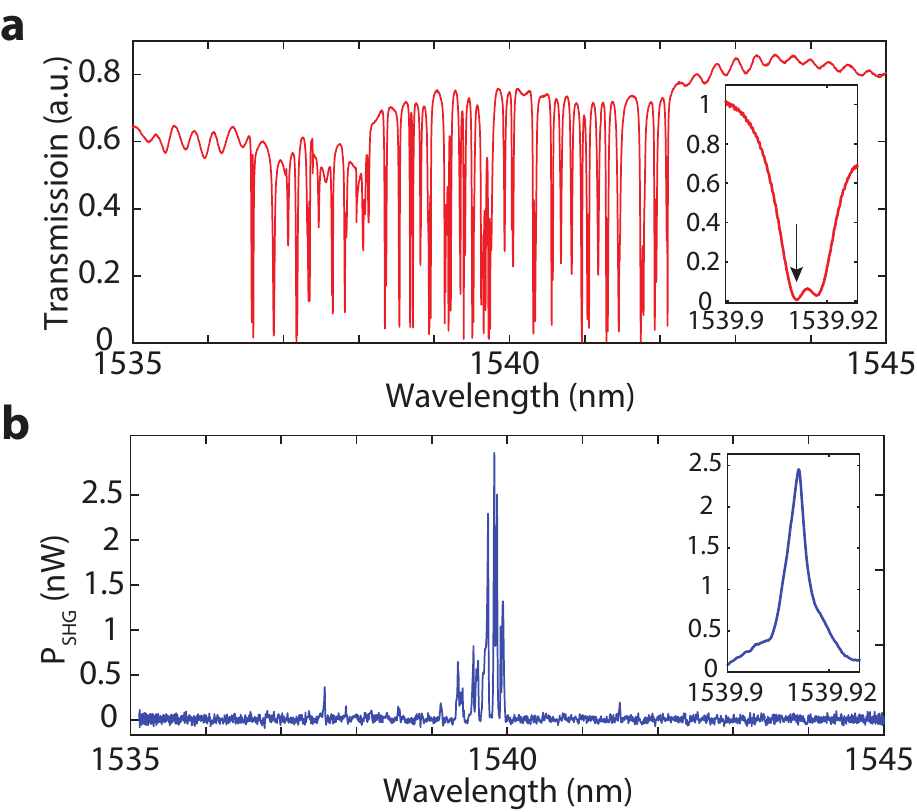}
\caption{\textbf{a}.  1550-nm band TE spectrum. Inset is the normalized spectrum of the phase- and frequency-matched resonance (arrow) used in the experiment. \textbf{b}. Second-harmonic power spectrum corresponding to the input pump wavelength (signal difference between the wide scan and the inset is due to the wavelength resolution of wide scan). }
\label{fig:spectrum}
\end{center}
\end{figure}

Fig. 2a shows scanning electron microscopy (SEM) images of the fabricated photonic circuit. The design of the phase-matched microring resonator follows Ref. \cite{zhao2022ingap}. The nominal ring radius is 5 $\mu$m. The microring resonator is coupled to two bus waveguides for transmitting the 1550-nm and 775-nm wavelength light, respectively. The 1550-nm band waveguide is 750 nm wide in the ring-waveguide coupled region and is separated from the ring by 350 nm. It decouples from the 775-nm TM$_{00}$ microring resonance because of the tight field confinement of the latter and the large ring-waveguide gap. On the other hand, the 775-nm wavelength pulley waveguide is 280 nm wide with a wrap angle of 6 degrees and a ring-waveguide gap of 250 nm. It decouples from the 1550-nm TE$_{00}$ microring resonance because of the significantly different mode index of the 1550-nm TE$_{00}$ mode in the narrow waveguide and the wide microring. Both bus waveguides are terminated with a photonic crystal mirror that is designed to reflect the 1550-nm TE and 775-nm TM light, respectively. At the end of the 1550-nm light waveguide, a section of 100-$\mu$m-long metal wire is fabricated on both sides of the waveguide in order to tune the phase of the waveguide via resistive heating and thermo-optic effect. The metal wires are 2 $\mu$m wide, 25 nm thick, and separated from the waveguide by 800 nm. The two bus waveguides are joined at a 1550-nm/775-nm wavelength-division multiplexer (WDM) which adiabatically couples the 1550-nm light between the two adjacent waveguides while forbidding the coupling of 775-nm light. One waveguide of the WDM is then connected to a tapered fiber-optic coupler which couples both 1550-nm TE polarized and 775-nm TM polarized light from a tapered fiber into the photonic circuit with an efficiency of about $60\%$ and $20\%$, respectively. The other waveguide of the WDM is tapered down to a width of 30 nm to reduce reflection from the waveguide end. An array of microrings with the width swept at 1 nm step are fabricated to achieve phase- and frequency-matched resonators. We also group 50 microrings in one device which effectively enhances the probability to realize the matching condition of a device and simplifies the measurement. The devices are electronically wired in parallel and can be tuned by a DC voltage. Fig. \ref{fig:spectrum}a shows the measured 1550-nm band reflection spectrum of a device.

\begin{figure}[!htb]
\begin{center}
\includegraphics[width=0.5\columnwidth]{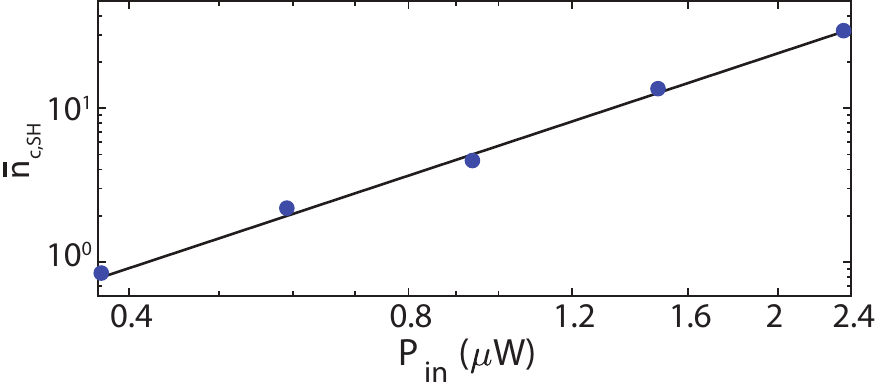}
\caption{Second-harmonic cavity photon number for varying input power at 1539.914 nm. Solid line is the model fitting.  }
\label{fig:shg}
\end{center}
\end{figure}

To characterize the fiber coupling efficiency, we fabricated on the same chip a simple waveguide device that has two identical waveguide couplers at the two ends and measured the transmission efficiency $\eta_{\mathrm{t}}$ to obtain the one-way coupling efficiency $\eta_{a(b), \mathrm{c}}=\sqrt{\eta_{a(b), \mathrm{t}}}=$ $60\%$ and $20\%$ for 1550-nm band TE and 775-nm band TM light, respectively. We also fabricated a waveguide device that has the same waveguide coupler on one end and a photonic crystal mirror on the other end. By measuring the reflection efficiency, and together with the measured coupler efficiency, we can get the mirror reflection efficiency $\eta_{a(b), \mathrm{m}}=$ $90\%$ and $2\%$ for 1550-nm band TE and 775-nm band TM light, respectively. Then, we measure the total reflection coefficient $\eta$ for the actual device used in the experiment, $\eta=\eta_{\mathrm{c}}^2\cdot\eta_{\mathrm{WDM}}^2\cdot\eta_{\mathrm{m}}$, where $\eta_{\mathrm{WDM}}$ is the transmission efficiency of the on-chip WDM. By this, $\eta_{a(b),\mathrm{WDM}}$ are found to be $90\%$ and $98\%$, respectively. After the tapered fiber tip is glued to the device, $\eta_{a(b), \mathrm{c}}$ drops due to the higher refractive index of the adhesive than that of the fiber. By measuring the change in the total reflection coefficient of the device, we find the coupler efficiency drops to $29\%$ and $1.2\%$ for 1550-nm band TE and 775-nm band TM light, respectively.

The nonlinear mode coupling of the phase-matched microring resonator is characterized using SHG (see Section \ref{App:data}). For this resonance, the measured second-harmonic cavity photon number $\bar n_{c,\textrm{SH}}$ for varying input power is plotted in Fig. \ref{fig:shg}, leading to an SHG efficiency of $105000\pm13360\%$/W. 

Table \ref{tab:deviceparameter} summarizes the key parameters for a typical phase- and frequency-matched device.

\begin{table}[htbp]
	\centering
    \caption{\textbf{Summary of measured device parameters.}}
    \small
    \bgroup
\def\arraystretch{1.5}
	\begin{tabular}{|c|c|}
		\hline
		Nonlinear mode interaction ($g/2\pi$) &  6.5 MHz     \\
		\hline
		1550-nm resonance intrinsic quality factor ($Q_{ai}$) &  $2.5\times 10^5$   \\
		\hline
		775-nm resonance intrinsic quality factor ($Q_{bi}$)    &  $5.1\times 10^4$     \\
		\hline
		775-nm resonance external quality factor ($Q_{be}$)   &   $2.0\times 10^6$ \\
		\hline
		1550-nm TE photonic crystal mirror reflection  & 90\%  \\
		\hline
		775-nm TM photonic crystal mirror reflection & 2\%  \\
		\hline
		Fiber coupler efficiency for 1550-nm TE & before glue: 60\%; after glue: 29\%  \\
		\hline
		Fiber coupler efficiency for 775-nm TM & before glue: 20\%; after glue: 1.2\%  \\
	    \hline
		Fiber coupler direct reflection for 1550-nm light & -30 dB \\
		\hline
		WDM efficiency for 1550-nm TE & 90\% \\
		\hline
		WDM efficiency for 775-nm TM & 98\%  \\
		\hline
		
	\end{tabular}%
	\egroup
	\label{tab:deviceparameter}
\end{table}

\subsection{On-chip wavelength-division multiplexer}

An on-chip WDM is used to (de)multiplex the 1550-nm and 775-nm band light, which are transmitted via individual waveguides for optimal coupling with the microring resonator. The on-chip WDM also enables the use of a single fiber-optic coupler for coupling of both 1550-nm and 775-nm band light between the optical fiber and the device. 

\begin{figure}[!htb]
\begin{center}
\includegraphics[width=0.7\columnwidth]{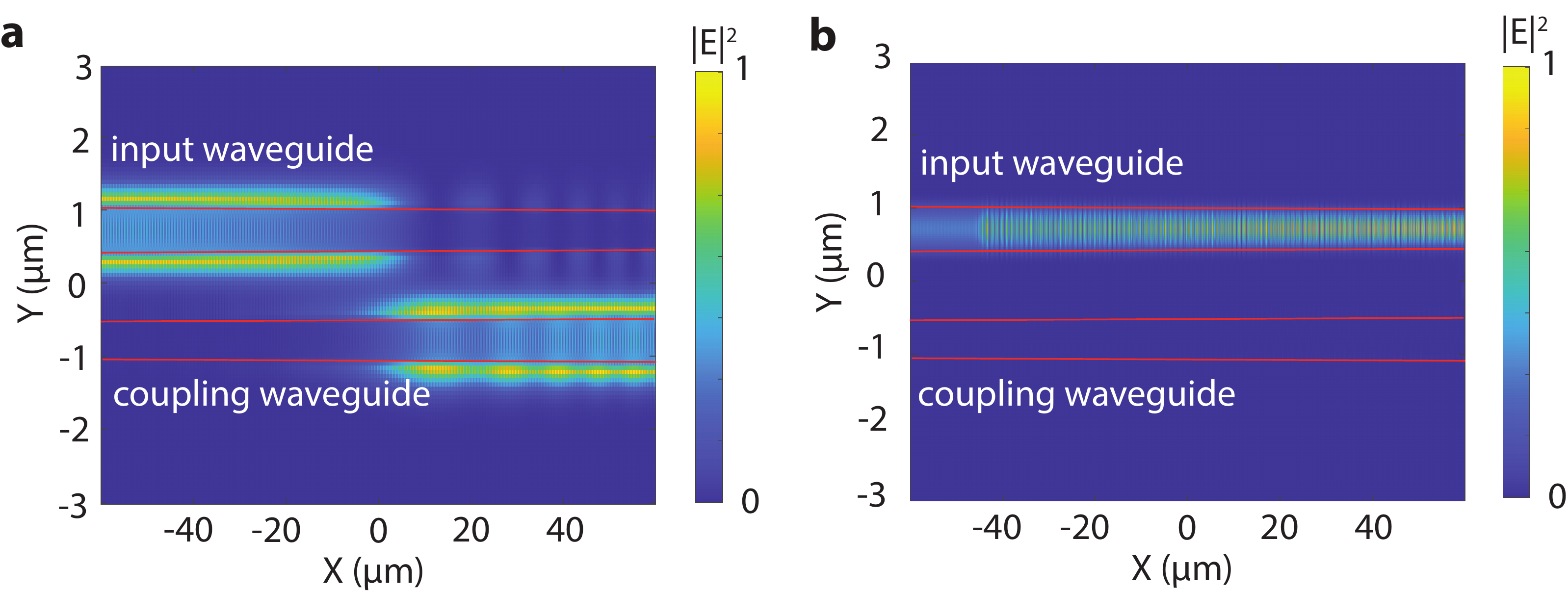}
\caption{FDTD simulation of propagation of 1550-nm TE$_{00}$ light (\textbf{a}) and 775-nm TM$_{00}$ light (\textbf{b}) in the WDM. }
\label{fig:wdm}
\end{center}
\end{figure}

The WDM consists of two adiabatically tapered waveguides in parallel. The gap between the two waveguides is sufficiently large such that the 775-nm-band TM$_{00}$ mode does not couple between the two waveguides. The input and coupling waveguide is adiabatically tapered down (from 430 nm to 290 nm over a length of 100 $\mu$m) and up (from 290 nm to 430 nm), respectively, such that the 1550-nm-band TE$_{00}$ light couples between the two adjacent waveguides with a high efficiency $>90\%$. Fig. \ref{fig:wdm} shows the finite-difference time-domain (FDTD) simulation of the transmission of the 1550-nm-band TE$_{00}$ light and 775-nm-band TM$_{00}$ light through the WDM.

\subsection{Resistive phase shifter}

A resistive heater is used to tune the waveguide phase. The phase change in the tuning section of the waveguide caused by the temperature change is given by 
\be
\Delta \phi=\Delta n\frac{\omega}{c}L=\frac{dn}{dT}\Delta T\frac{\omega}{c}L
\ee
where $L$ is the waveguide length and $\frac{dn}{dT}$ is the thermo-optic coefficient.

\begin{figure}[!htb]
\begin{center}
\includegraphics[width=0.7\columnwidth]{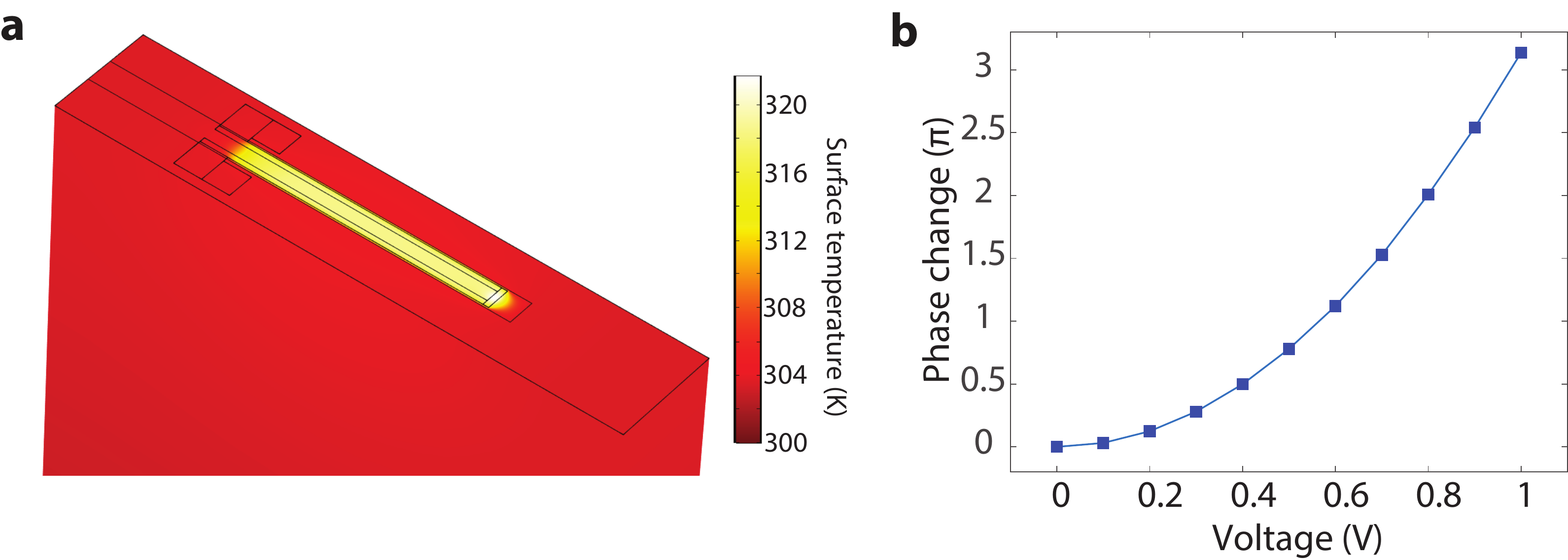}
\caption{\textbf{a}. Finite-element-method simulation of the temperature distribution around the release waveguide at 300 K for an applied voltage of 0.6 V. We used the room-temperature thermal-optic coefficient of InGaP, $2\times10^{-4}$ K$^{-1}$, and the room-temperature conductivity of gold, $4.517\times10^7$ S/m, in the simulation. \textbf{b}. Calculated waveguide phase change based on the simulated waveguide temperature for applied voltages. }
\label{fig:heater}
\end{center}
\end{figure}

In the device, the resistive heater is made from gold wires that are 100 $\mu$m long on both sides of the waveguide, 2 $\mu$m wide, 25 nm thick, and separated from the waveguide by 800 nm. Fig. \ref{fig:heater}a shows the temperature distribution near the waveguide/heater for a voltage of 0.6 V. Thanks to the released structure of the thin-membrane device, the heat is concentrated locally around the waveguide. Fig. \ref{fig:heater}b shows the calculated waveguide phase change versus applied voltage based on the simulated temperature of the waveguide.

Although the simulation is performed using the room-temperature material parameters, it turns out that the performance of the phase shifter at 4 K is close to the simulation. We observed that the waveguide phase can be tuned by $2\pi$ for a voltage $<$ 1 V.  Some literatures report significant drop of thermo-optic coefficient of materials at cryogenic temperatures; e.g., Ref. \cite{komma2012thermo} measured a four orders-of-magnitude reduction in bulk silicon substrate at 5 K compared to 300 K. Our measurement indicates that, however, the reduction of thermo-optic coefficient is likely to be mild and follows a similar reduction of thermal conductance for micro-/nano-structures in thin-film materials.

\section{Experimental setup}\label{App:setup}

\begin{figure}[!htb]
\begin{center}
\includegraphics[width=0.5\columnwidth]{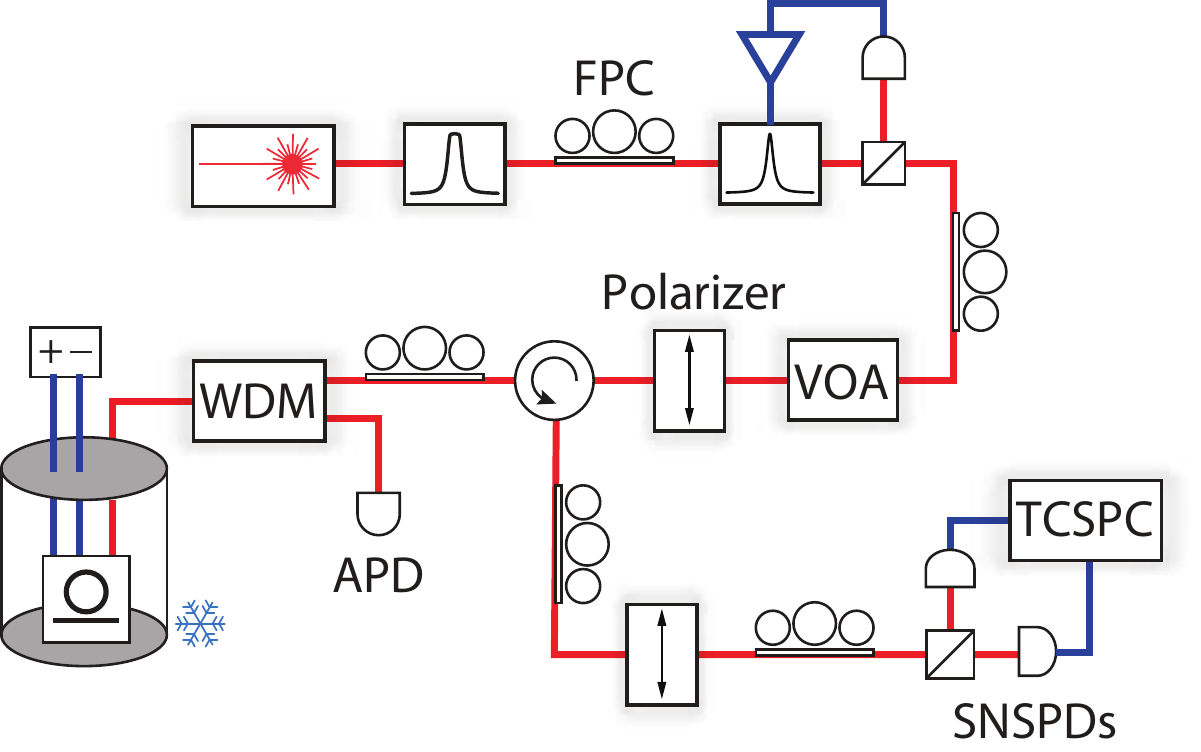}
\caption{\textbf{a}. Schematic of the experimental setup. FPC: fiber polarization controller. VOA: variable optical attenuator. WDM: wavelength-division multiplexer. APD: avalanche photodetector. SNSPDs: superconducting nanowire single-photon detector. TCSPC: time-correlated single-photon counting module. }
\label{fig:setup}
\end{center}
\end{figure}

Fig. \ref{fig:setup} shows a schematic plot of the experimental setup. Light from a tunable external cavity diode laser (New Focus TLB-6728) is filtered by an optical grating filter (JDS TB9223, 3 dB bandwidth 0.55 nm, 20 dB bandwidth 1.5 nm) and a fiber Febry-Perot (FP) filter (LUNA, 50 GHz free spectral range, 400 finesse) to eliminate the amplified spontaneous emission (ASE) and side-mode emission (at multiples of 2.4 GHz relative to the laser frequency) of the laser. One resonance of the FP filter is locked to the laser frequency via a dither feedback controller. 
The filtered light is then passed through a fiber polarizer (50 dB extinction) to eliminate unpolarized and orthogonally polarized light generated by random scattering in the fiber-optic circuit and, subsequently, a 1550-nm/775-nm WDM before coupling into the device. The WDM separates the 1550-nm and 775-nm (via second-harmonic generation) band light from the device. The 775-nm wavelength light is detected by an avalanche photodetector (Thorlabs APD440A). The 1550-nm wavelength light is further purified by a polarizer (40 dB extinction) and measured using the Hanbury-Brown and Twiss setup which consists of a 50:50 beamsplitter, two SNSPDs (Quantum Opus), and a time-correlated single-photon counting module (Time Tagger Ultra).

The device chip is wire-bonded to a printed circuit board for voltage tuning of the waveguide phase and is mounted in the mixing chamber of a dilution refrigerator (DR). The measurement, however, is performed at 4 K without condensation of the helium. A bare optical fiber is fed through the DR for transmitting the light. The tapered fiber tip is glued to the on-chip coupler for low-loss, mechanically-rigid fiber-optic coupling that is immune to the DR vibration (see Section \ref{App:cryo}).

\section{Noise sources and mitigation}\label{App:noise}

\subsection{Amplified spontaneous emission, laser side modes, and unpolarized light}

In our experiment, tuning of the transmission of the photonic circuit near the critical-coupling regime relies on the destructive interference of light. Any incoherent light will lead to residual transmission which might obscure the signal. The major incoherent light sources in the experimental setup include the amplified spontaneous emission of the laser, laser side-mode emissions, and unpolarized light. To eliminate the broadband amplified spontaneous emission, we used a wavelength-tunable optical grating filter (JDS TB9223, 3 dB bandwidth 0.55 nm, 20 dB bandwidth 1.5 nm), which lowered the minimum normalized transmission from $2\times 10^{-3}$ to $1\times 10^{-4}$ at room temperature. We also used two polarizers (40 dB and 50 dB extinction ratio) to eliminate the unpolarized and random polarized light from the laser or generated through scattering in the optical fiber setup. At this level, the minimum transmission is limited by the thermomechanical and thermorefractive noises from the device (see below). To suppress these thermal noises, we placed the device chip in a dilution refrigerator but operated at 4 K. With the same setup, the minimum transmission is reduced to $1\times 10^{-5}$, which is nevertheless higher than the residual thermal noise supposed to be at 4 K. We further used a narrow bandwidth fiber Fabry-Perot filter (LUNA, 50 GHz free spectral range, 400 finesse) to eliminate the laser side-mode emission at multiples of 2.4 GHz relative to the laser frequency as well as the residual ASE in the vicinity of the laser frequency. By doing so, the minimum transmission is reduced from $1\times 10^{-5}$ to $1\times 10^{-6}$, which is dominated by the thermomechanical noise at 4 K.

\subsection{Thermomechanical and thermorefractive noises}

Besides the external incoherent light which causes residual transmission, the photonic device itself also generates noise photons upon incidence of the laser light. The InGaP photonic circuit is suspended with a thin Al$_2$O$_3$ membrane, which supports vibration modes with frequencies in the range of a few MHz to a few hundred MHz that couple to the microring resonator via the radiation-pressure force \cite{zhao2022observation}. At finite temperatures, thermal motion of these vibration modes modulates the input light, generating Stokes and anti-Stokes (thermal) light, which cannot interfere with the input coherent signal. In addition, the phonon-scattered light has a super-Poissonian photon statistics with $g^{(2)}(0)>1$. At room temperature and for the 5-$\mu$m-radius ring, the equivalent transmission corresponding to the thermomechanical noise is about $10^{-4}$ \cite{zhao2022observation}. Thermomechanical noise is proportional to the phonon occupation and thus the temperature $T$.

Another source of thermal noise is the thermorefractive noise. The intrinsic fluctuation of the temperature in a finite volume causes fluctuation of the refractive index, leading to jitter of the cavity frequency. The cavity frequency jitter effectively modulates the input light and generates noise photons. At room temperature and for the 5-$\mu$m-radius ring, the equivalent transmission corresponding to the thermorefractive noise is about $10^{-4}$ \cite{zhao2022observation}. Thermorefractive noise has a quadratic dependence on the temperature ($\propto T^2$) \cite{gorodetsky2004fundamental,zhao2022observation}.

Since both thermal noises are temperature dependent, they can be suppressed at cryogenic temperatures. At 4 K, the equivalent transmission corresponding to the thermomechanical and thermorefractive noise is below $10^{-6}$ and $10^{-8}$, respectively. 

Another method to suppress thermal noises is by adding the top oxide cladding of the microring. The the optomechanical coupling between the microring and the membrane vibration mode is roughly proportional to $1/\sqrt{V}$ \cite{zhao2022observation}, where $V$ is the volume of the oxide cladding on the microring. Thus, the thermomechanical noise is proportional to $1/V$. For example, we find via finite element simulation that the optomechanical coupling reduced by a factor of 37x by adding 1 $\mu$m silicon dioxide cladding.  Thus, by adding a few micrometers of silicon dioxide cladding, e.g., via the PECVD process, one can suppress the thermomechanical noise by at least three orders of magnitude. Adding top oxide cladding also suppresses the temperature fluctuation given the larger mode volume \cite{gorodetsky2004fundamental}, leading to less thermorefractive noise.

Table \ref{tab:noise} summarizes major noise sources in our experiment and methods for mitigation.

\begin{table}[htbp]
	\centering
    \caption{\textbf{Summary of noise sources and methods of mitigation (values are for $R=5~ \mu$m ring ).}}
    \small
    \bgroup
\def\arraystretch{1.5}
	\begin{tabular}{|c|ccc|}
		\hline
		Noise & Mitigation     & Before    & After    \\
		\hline
		Unpolarized light    & Polarizer     &  $3\times10^{-3}$    &  $2\times10^{-3}$ \\
		\hline
		ASE & Filter     & $2\times10^{-3}$    & $10^{-5}$      \\
		\hline
		Laser side mode & Filter & $10^{-5}$   & $<10^{-6}$  \\
		\hline
		Thermomechanical noise &  4 K/thick oxide cladding & $10^{-4}$ & $10^{-6}/10^{-7}$  \\
		\hline
		Thermorefractive noise &  4 K/thick oxide cladding & $10^{-4}$ & $10^{-8}/10^{-5}$   \\
		\hline
	\end{tabular}%
	    \egroup
	\label{tab:noise}
\end{table}

\section{Cryogenic nonlinear photonics experiment}\label{App:cryo}
 
We outline the procedure of performing the nonlinear photonics experiment in a dilution refrigerator (DR), including identifying the phase-matched device and realizing mechanically-stable fiber-optic coupling which is particularly important for this experiment.

\subsection{Frequency- and phase-matched device}
\begin{figure}[!htb]
\begin{center}
\includegraphics[width=0.8\columnwidth]{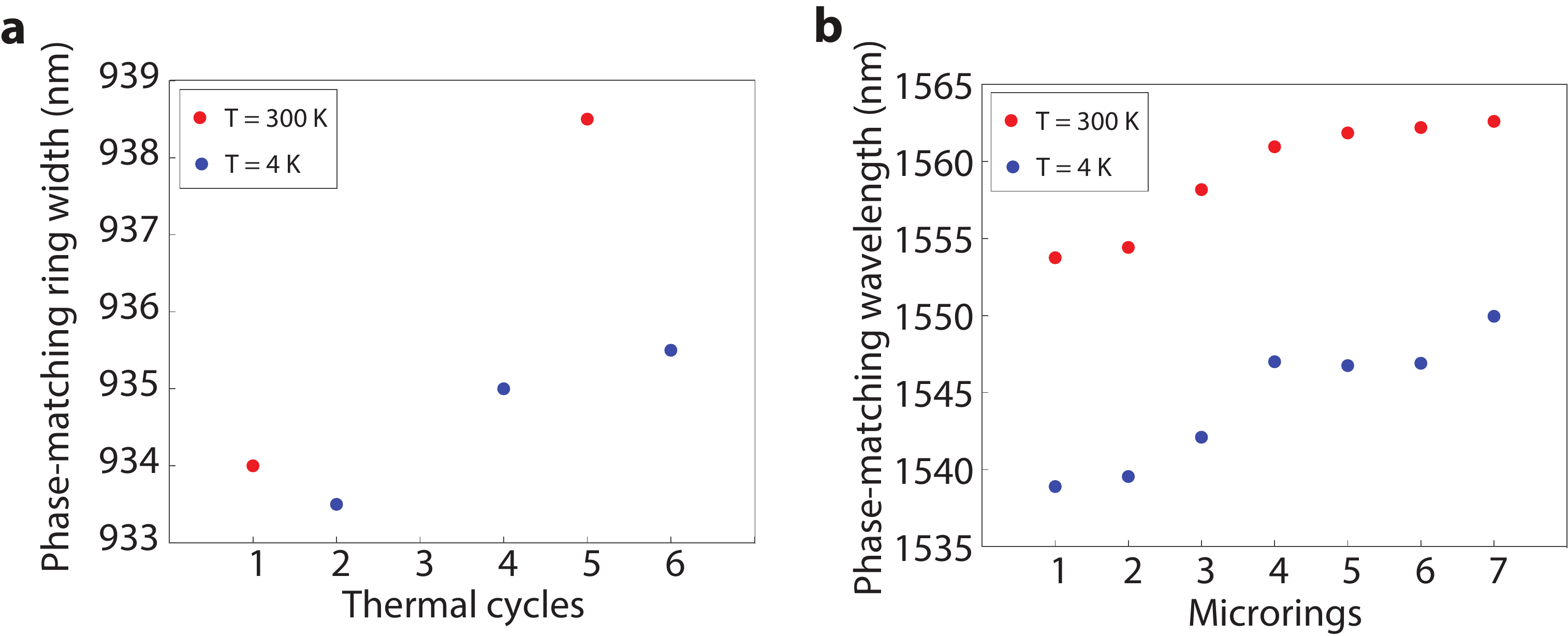}
\caption{\textbf{a}. The change of phase-matching ring width of one chip during multiple temperature cycles. \textbf{b}. The shift of the 1550 nm band phase-matching resonance wavelength in multiple chips from 300 K to 4 K.}
\label{fig:Tcycle}
\end{center}
\end{figure}

Finding the frequency- and phase-matched device is achieved by scanning microrings with different width and measuring the SHG efficiency, as the phase-matched ring yields the greatest SHG efficiency. A tapered fiber is installed in the DR and the chip is mounted on a piezostage for scanning devices. The DR has an optical window which allows imaging and alignment of the device and tapered fiber. However, device scanning in the DR could be time-consuming. To mitigate this, we first perform the measurement in a room-temperature setup and identify the phase-matched device. Since the frequency shift of the 1550-nm and 775-nm band resonances from room temperature to 4 K is different, the phase-matched device at 4 K will be different from the one at room temperature. However, we find such a change, despite varying for each temperature cycle and each chip, is only a few devices corresponding to the change of ring width by a few nm.  Fig. \ref{fig:Tcycle} shows the change of phase-matching ring width measured for several temperature cycles and change of the 1550-nm resonance wavelength of the phase-matched ring. To further simplify the measurement, we fabricated multiple microring resonators into a single device, so they can be measured altogether at once. Effectively, it enhances the probability for a single device to satisfy the phase- and frequency-matching condition.  Such device arrangement also mitigates the possible cavity frequency discontinuity caused by the thickness nonuniformity of the device layer across large areas.

\subsection{Mechanically-rigid fiber-optic coupler}
Another challenge associated with the fiber-optic measurement in a DR is the vibration. The vibration of the DR causes relative movement between the tapered fiber and the chip even when they are in touch, leading to fluctuations of both the transmission and direct reflection of the fiber-optic coupler. Since the device is operated near the critical-coupling regime, signal transmission is sensitive to such fluctuations. To overcome this, after finding the phase-matched device at low temperature, we glue the tapered fiber to the device to make a mechanically-rigid fiber-optic coupling. 

We first glued the tapered fiber tip to the tapered waveguide. We used a UV-cured, tack-free optical adhesive (Norland NOA 61, refractive index 1.56), as it allows a thin layer around the fiber tip and is 100$\%$ solid after curing. The coupling efficiency is optimized by adjusting the tip position before UV curing. Because the applied glue is only a very thin layer, the coupling efficiency of the adiabatic coupler remains high (about $30\%$ for the 1550 nm-band light).  Next we applied a large bead of glue on the edge of the chip to fix the fiber and let the glue to flow by a few mm along the fiber before UV curing. For this, we used medium viscosity, low outgassing optical adhesive (Norland NOA 88, refractive index 1.56). Finally, the chip is heated to 50 C for 12 hours for aging of the glue to achieve the optimum adhesion.

\section{Data analysis}\label{App:data}

\subsection{Nonlinear mode coupling}
For the microring cavity that supports a pair of phase- and frequency-matched fundamental and second-harmonic resonances with frequency $\omega_a$ and $\omega_b$, respectively, the interaction Hamiltonian between the two resonances is given by 
\be
\hat H_{\textrm{int}}= \hbar g(\hat a^{\dagger 2}\hat b+\hat a^2\hat b^\dagger).
\ee
The nonlinear coupling $g$ can be determined by the SHG efficiency. In the pump non-depletion region, the normalized SHG efficiency, $\eta  \equiv \frac{P_\mathrm{SHG}}{P_\mathrm{in}^2}$, where $P_\mathrm{SHG}$ and $P_\mathrm{in}$ are the on-chip SHG and pump power, respectively, is given by \cite{guo2016second}
\begin{equation}
	\eta=g^2\frac{\kappa_{b,e}/2}{\Delta_{b}^2+(\kappa_{b}/2)^2}\bigg(\frac{\kappa_{a,e}}{\Delta_{a}^2+(\kappa_{a}/2)^2}{}\bigg)^2\frac{\hbar\omega_{b}}{(\hbar\omega_{a})^2},
\label{eqn:SHG}
\end{equation}
where $\kappa_{a(b)}\equiv \kappa_{a(b),e}+\kappa_{a(b),i}$, $\kappa_{a(b),e}$  and $\kappa_{a(b),i}$ are the total, external and intrinsic photon (energy) loss rate of the resonances, respectively, $\Delta_{a}=\omega_{a}-\omega_p$, $\Delta_{b}=\omega_{b}-2\omega_p$, and $\omega_p$ is the frequency of the pump. We have used the fact that the fundamental 1550 nm-band resonance is split standing-wave resonances and they couple to the even 775 nm-band resonance, regardless whether the latter is split or not. The 775 nm-band photonic crystal mirror has a weak reflection and thus only one direction of the emitted SHG light is detected. The second-harmonic cavity photon number is given by 
\bqa\nonumber
\bar n_{c,\mathrm{SH}}&=&\frac{g^2}{\Delta_{b}^2+(\kappa_{b}/2)^2}\bigg(\frac{\kappa_{a,e}/2}{\Delta_{a}^2+(\kappa_{a}/2)^2}{}\bigg)^2\frac{P^2_{\mathrm{in}}}{(\hbar\omega_{a})^2}\\
&=&\frac{g^2\bar n_c^2}{\Delta_{b}^2+(\kappa_{b}/2)^2},
\eqa
where $\bar n_c$ is the cavity photon number (of the pump).

The phase-matched resonance at 1539.914 nm is tuned close to the critical-coupling condition when performing the SHG measurement. At this condition, the split between the two transmission dips is much smaller than the cavity linewidth, and thus we used the single-resonance transmission coefficient formula $|t_{\omega}|^2=|\frac{(\kappa_e-\kappa_i)/2+i(\omega-\omega_0)}{(\kappa_e+\kappa_i)/2-i(\omega-\omega_0)}|^2$ to fit the resonance transmission spectrum. The internal and external quality factors are extracted to be $Q_{a,i}\approx Q_{a,e}\approx 2.5\times 10^5$. The 775-nm TM$_{00}$ resonance is designed to be highly under-coupled, and we find $Q_{b,i}\approx 5.0\times 10^4$ and $Q_{b,e}\approx 2.0\times 10^6$. Based on the measured circuit efficiency (Section \ref{App:PIC}) and the SHG efficiency of $105000\pm 13360$\%/W, the nonlinear mode coupling rate is then found to be $g/2\pi=6.5$ MHz. This value is consistent with the coupling rate measured in Ref. \cite{zhao2022ingap}, wherein the coupling rate is referred to that of the non-split resonances which differs by a factor of $\sqrt{2}$ from the split one in theory.

\subsection{Tuning minimum transmission}
The minimum transmission coefficient for waveguide phase $\varphi$ has the relation $t_{\mathrm{min}}\propto \Delta\varphi=\varphi-\varphi_0$, where $\varphi_0$ corresponds to $t_{\mathrm{min}}\approx 0$. Since $\Delta\varphi\propto \Delta T \propto \Delta (V^2)$, we have $t_{\mathrm{min}}^2\propto (\Delta (V^2))^2$, where $\Delta T$ is the temperature increase relative to $T_0$ at the phase shifter and $V$ is the voltage applied to the phase shifter. The data in Fig. 2g is fitted using $t_{\mathrm{min}}^2=\lambda (V^2-V_0^2)^2$, where $\lambda$ and $V_0$ are fitting parameters. 

\subsection{Second-order correlation function}

\begin{figure}[!htb]
\begin{center}
\includegraphics[width=1\columnwidth]{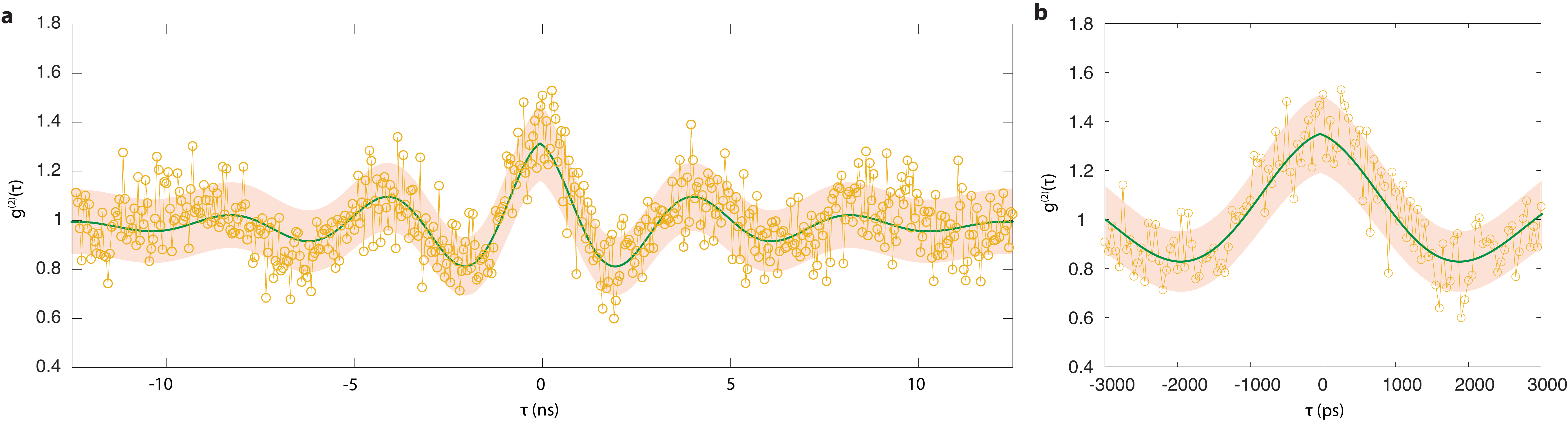}
\caption{\textbf{a}. Measured $g^{(2)}(\tau)$ of a phase-unmatched resonance, showing the thermo-mechanical noise. \textbf{b}. Zoom-in view of \textbf{a}.}
\label{fig:unmatch}
\end{center}
\end{figure}

Due to the presence of the residual thermomechanical noise, the output light from the device is a combination of coherent transported photons and thermomechanical noise \cite{zhao2022observation}, which is expressed as
\begin{equation}
a_{\mathrm{out}}=t_\omega a_{s}+ua_{m},
\end{equation}
where $a_{s}$ and $a_{m}$ are the annihilation operators corresponding to the transported signal and thermomechanical noise, respectively, $t_\omega$ is the transmission coefficient, and $u$ is the relative amplitude of the thermomechanical noise. The measured second-order correlation function of the total output state is given by
\begin{widetext}
\begin{equation}\label{eq:average model}
\begin{aligned}
g^{(2)}(\tau)&=\frac{\langle a_{\rm{out}}^{\dag}(0)a_{\rm{out}}^{\dag}(\tau)a_{\rm{out}}(\tau)a_{\rm{out}}(0) \rangle}{\langle a_{\rm{out}}^{\dag}(0)a_{\rm{out}}(0)  \rangle ^2}\\
&=\frac{|t_\omega|^4 g_{s}^{(2)}(\tau)+|u|^4 g_{m}^{(2)}(\tau)+2|t_\omega u|^2g_{s}^{(1)}(\tau)g_{m}^{(1)}(\tau) \cos(\omega_m\tau)+2|t_\omega u|^2g_{s}^{(1)}(0)g_{m}^{(1)}(0)}{(|t_\omega|^2+|u|^2)^2}\\
&=(g_s^{(2)}(\tau)-1)\frac{|t_\omega|^4}{(|t_\omega|^2+|u|^2)^2}+(g_m^{(2)}(\tau)-1)\frac{|u|^4}{(|t_\omega|^2+|u|^2)^2}+\frac{2 |t_\omega u|^2}{(|t_\omega|^2+|u|^2)^2}g_m^{(1)}(\tau)\cos(\omega_m\tau)+1,
\end{aligned}
\end{equation}
\end{widetext}
where $g_{s(m)}^{(1)}(\tau)$ and $g_{s(m)}^{(2)}(\tau)$ are the first-order and second-order correlation functions of the signal(thermomechanical noise), respectively, and $\omega_m$ is the frequency of the radiation-pressure-coupled mechanical mode.  We have used $g_s^{(1)}(\tau)=1$, $g_m^{(1)}(\tau)=e^{-\gamma_m\tau/2}$, and $g_m^{(2)}(\tau)=1+e^{-\gamma_m\tau}$, where $\gamma_m$ is the dissipation rate of mechanical mode. We only keep one mechanical mode in the fitting because the dominant thermomechanical noise is originated from the ring breathing mode whose frequency is about 250 MHz according the simulation. The thermomechanical noise is existent for both phase-matched and unmatched microrings. Fig. \ref{fig:unmatch} shows the measured $g^{(2)}$ for a phase-unmatched microring. The oscillation of $g^{(2)}(\tau)$ matches the frequency of the breathing mode.

Since the correlation $g_{s}^{(2)}(\tau)$ of the signal due to the photon-photon interaction in the $\chi^{(2)}$ cavity has a much shorter characteristic time scale ($\sim 1/\kappa_a$) compared to the period of the mechanical mode, we can use the last two terms in Eq. \ref{eq:average model} to fit the measured $g^{(2)}(\tau)$ for $\tau\gg 1/\kappa_a$ to obtain the ratio $|u/t_\omega|$. Then we use this information and the full Eq. \ref{eq:average model} to fit the whole $g^{(2)}(\tau)$ using Eq. \ref{fitg2}-\ref{fitT}. We assumed $\omega_b=2\omega_a$ and used global fitting to fit the measured $g^{(2)}(\tau)$ at different transmission level $|t_\omega|^2$, by treating $\kappa_{a,i}$ as a global fitting parameter while $\delta$, $\Delta$, $\kappa_{a,e}$ as variable fitting parameters.

